\documentclass[aps,prl,twocolumn,superscriptaddress,showpacs,reprint]{revtex4-1}

\usepackage[ansinew]{inputenc}
\usepackage[T1]{fontenc}
\usepackage{ae,aecompl}
\usepackage[english]{babel}
\usepackage{graphicx}
\usepackage{hyperref}
\usepackage{epstopdf}
\usepackage{color}
\usepackage{multirow}
\usepackage{natbib}

\begin{document}

\author{F. Arnold}
\affiliation{Max Planck Institute for Chemical Physics of Solids, 01187 Dresden, Germany}
\author{M. Naumann}
\affiliation{Max Planck Institute for Chemical Physics of Solids, 01187 Dresden, Germany}
\affiliation{Physik-Department, Technische Universit\"at M\"unchen, 85748 Garching, Germany}
\author{S. Khim}
\affiliation{Max Planck Institute for Chemical Physics of Solids, 01187 Dresden, Germany}
\author{H. Rosner}
\affiliation{Max Planck Institute for Chemical Physics of Solids, 01187 Dresden, Germany}
\author{V. Sunko}
\affiliation{Max Planck Institute for Chemical Physics of Solids, 01187 Dresden, Germany}
\affiliation{Scottish Universities Physics Alliance, School of Physics and Astronomy, University of St. Andrews, St. Andrews, Fife KY16 9SS, UK}
\author{F. Mazzola}
\affiliation{Scottish Universities Physics Alliance, School of Physics and Astronomy, University of St. Andrews, St. Andrews, Fife KY16 9SS, UK}
\author{P.D.C. King}
\affiliation{Scottish Universities Physics Alliance, School of Physics and Astronomy, University of St. Andrews, St. Andrews, Fife KY16 9SS, UK}
\author{A.P. Mackenzie}
\thanks{correspondence should be addressed to mackenzie@cpfs.mpg.de and elena.hassinger@cpfs.mpg.de}
\affiliation{Max Planck Institute for Chemical Physics of Solids, 01187 Dresden, Germany}
\affiliation{Scottish Universities Physics Alliance, School of Physics and Astronomy, University of St. Andrews, St. Andrews, Fife KY16 9SS, UK}
\author{E. Hassinger}
\thanks{correspondence should be addressed to mackenzie@cpfs.mpg.de and elena.hassinger@cpfs.mpg.de}
\affiliation{Max Planck Institute for Chemical Physics of Solids, 01187 Dresden, Germany}
\affiliation{Physik-Department, Technische Universit\"at M\"unchen, 85748 Garching, Germany}

\title{Quasi two-dimensional Fermi surface topography of the delafossite PdRhO$_2$}

\date{\today}

\begin{abstract}
We report on a combined study of the de Haas-van Alphen effect and angle resolved photoemission spectroscopy on single crystals of the metallic delafossite PdRhO$_2$ rounded off by \textit{ab initio} band structure calculations. A high sensitivity torque magnetometry setup with SQUID readout and synchrotron-based photoemission with a light spot size of $~50\,\mu\mathrm{m}$ enabled high resolution data to be obtained from samples as small as $150\times100\times20\,(\mu\mathrm{m})^3$. The Fermi surface shape is nearly cylindrical with a rounded hexagonal cross section enclosing a Luttinger volume of 1.00(1) electrons per formula unit.
\end{abstract}

\maketitle

\noindent In recent years delafossite layered metallic oxides \cite{Shannon71} have attracted considerable attention because of their extremely high electrical conductivity and the simplicity of their electronic structure \cite{Mackenzie17}. The delafossite structure of general formula ABO$_2$ features alternating triangularly co-ordinated A metal layers separated by BO$_2$ layers in which B is a transition metal in a trigonally distorted octahedral co-ordination with oxygen \cite{Prewitt71}. The layer stacking sequence results in there being three formula units per hexagonal unit cell, with the space-group $R\overline{3}m$. Many delafossites are semiconducting or insulating, but those with A site metals Pd or Pt are highly anisotropic metals in which conductivity in the layers is hundreds of times larger that that perpendicular to them. Even at room temperature, the in-plane resistivities of non-magnetic PtCoO$_2$ and PdCoO$_2$ are just over $2\,\mu\Omega\mathrm{cm}$ \cite{Hicks12,Kushwaha15}, lower than that of any elemental metal except Ag and Cu. Taking into account the factor of three lower carrier density in the delafossites, they have a room temperature mean free path at least a factor of two longer than even that of pure Ag. The resistivity falls rapidly with temperature, and resistive mean free paths of over $20\,\mu\mathrm{m}$ have been observed in PdCoO$_2$ \cite{Hicks12}.

The Fermi surface of the known delafossite metals is extremely simple.  In non-magnetic PdCoO$_2$ and PtCoO$_2$, it is a single, weakly corrugated cylinder with nearly hexagonal cross-section \cite{Kushwaha15,Eyert08,Kim09,Ong10,Noh09}. In PdCrO$_2$, a similar cylinder is observed above $40\,\mathrm{K}$, but at low temperatures very small gapping is detected, due to coupling between spin ordering in the CrO$_2$ layers and the states in the broad conduction band whose dominant character is Pd $4d/5s$-like \cite{Sobota13,Ok13,Noh14,Hicks15}. Electron counting in PdCrO$_2$ highlights the role of correlations in the transition metal layer of the delafossites: the CrO$_2$ layer is Mott insulating \cite{Hicks15}.  

The knowledge to date of the delafossite metals therefore points to an interesting and very unusual situation in which there is a close interplay between an extremely broad conduction band with a Fermi velocity of order $8\times10^5\,\mathrm{ms}^{-1}$ (close to the free electron value) and $3d$ transition metal states for which correlations are known to be strong.  The situation is made even richer by the fact that the weakly- and strongly-correlated states arise from different layers in the crystal structure. Delafossites are like a naturally-occurring example of the kind of heterostructures that many groups world-wide are trying to synthesize artificially, and a natural structural class on which to base future layer-by-layer synthesis.

The unique combination of properties highlighted above has already led to the observation of fascinating physics, notably the observation of huge c-axis magnetoresistance oscillations \cite{Takatsu13,Kikugawa16}, the unconventional Hall effect \cite{Takatsu10}, and hydrodynamic electron flow \cite{Moll16}, and it seems likely that new regimes of mesoscopic transport will be attainable via focused ion beam microstructuring of single crystals. 

All of these phenomena are expected to be strongly sensitive to the details of the Fermi surface shape, i.e. the curvature of the in-plane hexagon, as well as the out-of-plane warping. To unlock the full potential of the delafossite oxides and to yield new physics, it is crucial to have access to slightly different Fermi surface topographies and different levels of correlation in the ABO$_2$ layers, while preserving the overall simplicity of the electronic structure. There is a pressing need, therefore, to have as many such metals available for precision study as possible.  So far, the only monovalent delafossite metals for which single crystals exist are PdCoO$_2$, PdCrO$_2$ and PtCoO$_2$ in which the B-site cations are $3d$ transition metals \cite{Coldea09}. Based on preliminary studies on powders and polycrystalline thin films, as well as electronic structure calculations \cite{Baird88,Carcia80,Kim14}, PdRhO$_2$ is thought to be metallic and also to have a single conduction band. Hence this material offers the opportunity to study the effect of varying Pd-Pd overlap integrals, as well as the effect of changing on-site correlation and spin-orbit coupling strengths by moving to a $4d$ B site transition metal.

Recently, we have succeeded in crystallizing PdRhO$_2$ \cite{Kushwaha17}. Here, we report a comprehensive study of de Haas-van Alphen (dHvA) measurements on this new material, and combine the dHvA data with information from angle resolved photoemission spectroscopy (ARPES) to determine the Fermi surface with high precision. We also highlight the potential of PdRhO$_2$ to test and refine the accuracy of modern many-body electronic structure calculations. 

\begin{figure}[t]
	\centering
		\includegraphics[width=\columnwidth]{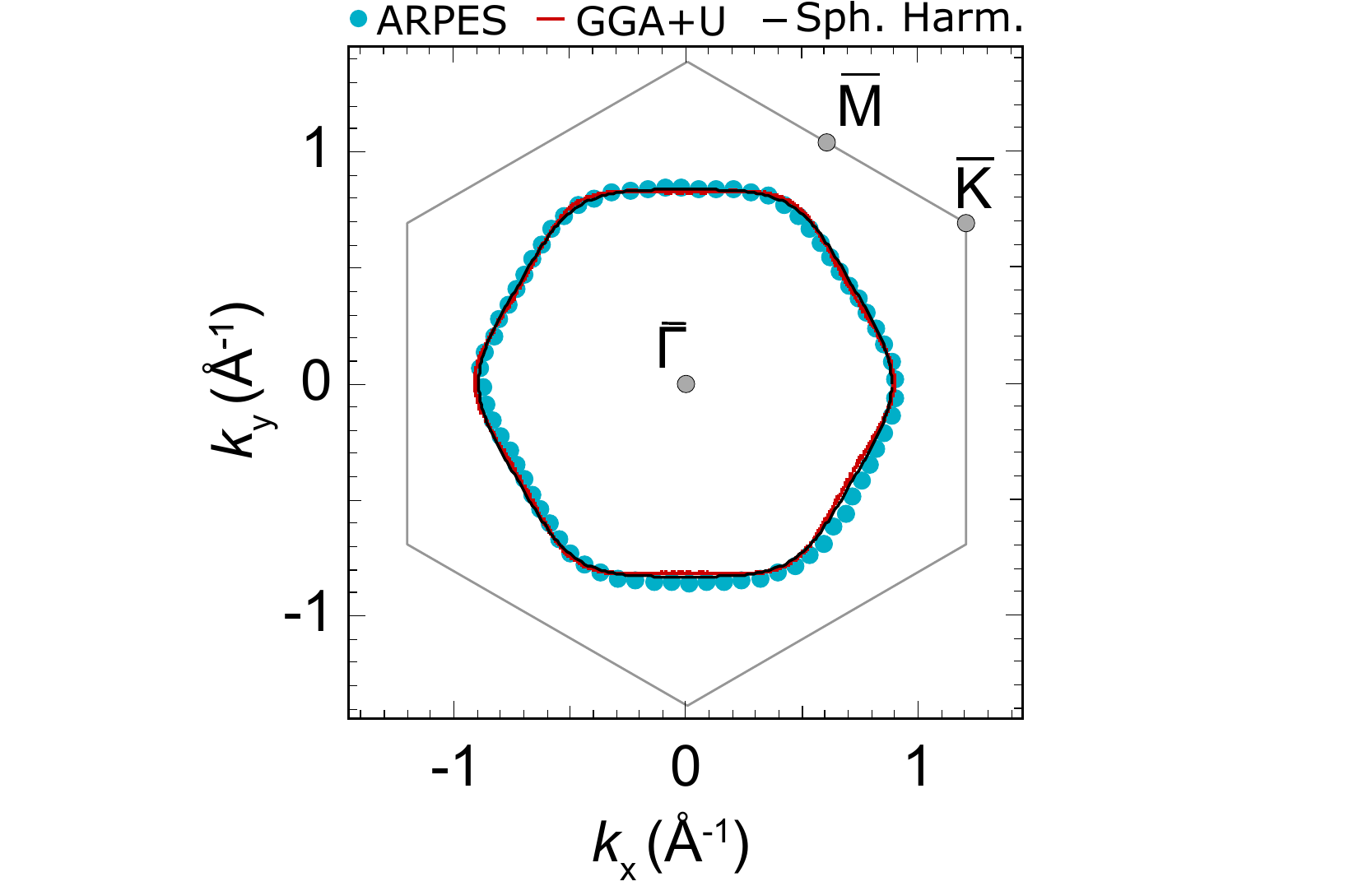}
	\caption{Planar Fermi surface topography of PdRhO$_2$ as determined by angle-resolved photo-emission spectroscopy (ARPES), density functional theory (DFT) calculations, and in-plane Fermi surface harmonics quoted later in Tab.~\ref{tab:SphHamExp}. DFT and surface harmonics data have been rescaled by $-6\,\%$ to fit the ARPES results.}
	\label{fig:ARPES-FS}
\end{figure}

Crystal growth and characterization of single crystals of PdRhO$_2$ is described in \cite{Kushwaha17,SOM}. De Haas-van Alphen oscillations of two PdRhO$_2$ crystals from the same growth batch were observed at temperatures between $100\,\mathrm{mK}$ and $4\,\mathrm{K}$ in magnetic fields up to $15\,\mathrm{T}$. The respective sample sizes were approximately $200\times300\times50\,(\mu\mathrm{m})^3$ and $150\times100\times20\,(\mu\mathrm{m})^3$.
Experiments were performed using an ultra-low noise SQUID torque magnetometer, installed on a MX400 Oxford Instruments dilution refrigerator with a $15/17\,\mathrm{T}$ superconducting magnet and $270^o$ Swedish rotator with an angular accuracy of $\Delta\theta=\pm0.2^\circ$. The magnetometer utilizes piezoresistive PRC400 micro-cantilevers and a two-stage dc-SQUID as highly sensitive read-out, offering an unprecedented torque resolution of $\Delta\tau=2\times10^{-13}\,\mathrm{Nm}$ at lowest temperatures \cite{Arnold17RSI,Rossel96}. Data were taken at constant temperatures whilst the magnetic field was swept from 15 to $7.5\,\mathrm{T}$ at a rate of $30\,\mathrm{mT/min}$. 

ARPES was performed using the I05 beamline of Diamond Light Source, UK. Samples were cleaved in-situ at the measurement temperature of $13\,\mathrm{K}$, and probed using linear horizontal polarisation light with a photon energy of $110\,\mathrm{eV}$ and spot size of $\approx50\,\mu\mathrm{m}$. As well as the bulk Fermi surface extracted here, surface states indicative of a RhO$_2$-termination were also observed in the experiment \cite{Sunko17}.

Relativistic density functional (DFT) electronic structure calculations including spin-orbit coupling were performed using the full-potential FPLO code \cite{Koepernik99,Opahle99,FPLO}, version fplo14.00-47 within the general gradient approximation (GGA). Coulomb repulsion in the Rh-$4d$ shell was simulated in a mean field way applying the GGA+$U$ approximation in the atomic-limit-flavor \cite{Kushwaha17,SOM}. 

\begin{figure*}[t]
\centering
		\includegraphics[width=\textwidth]{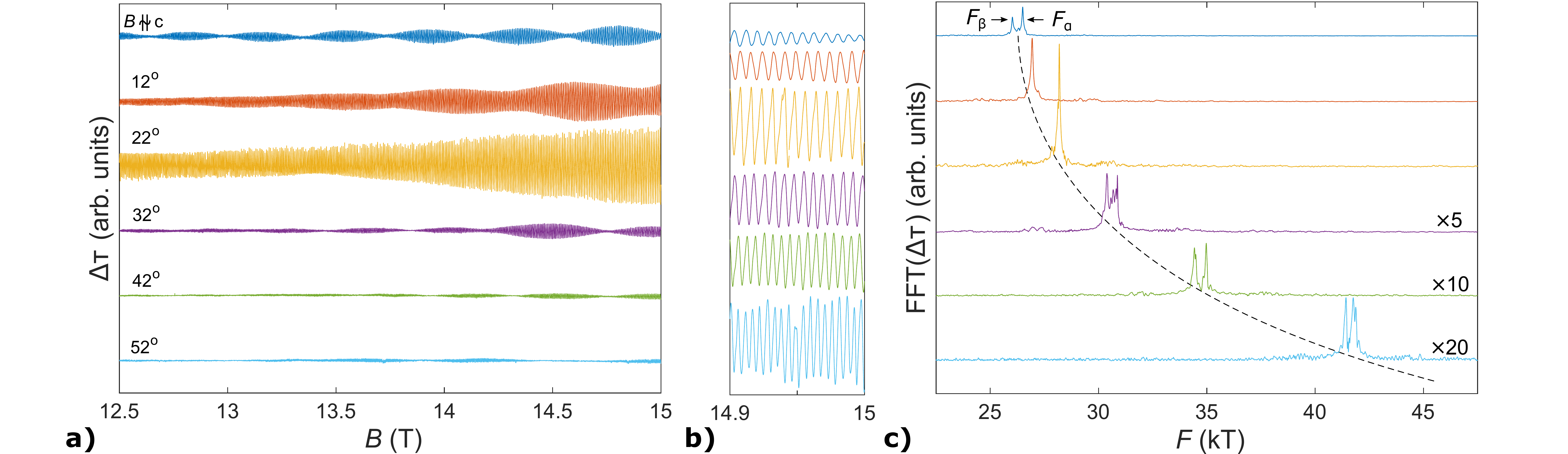}
\caption{Magnetic torque de Haas-van Alphen signal of PdRhO$_2$ at $T\approx100\,\mathrm{mK}$. \textbf{a)} shows the magnetic field dependence of the de Haas-van Alphen oscillations for a selection of magnetic field angles within the Z$\Gamma$L-plane. The data were background subtracted by a $2^\mathrm{nd}$-order polynomial. A zoom of the high field oscillations is shown in \textbf{b)}. \textbf{c)} displays the Fourier transforms corresponding to the oscillations shown in a). Data have been multiplied and offset for clarity. The dashed line shows the $1/\cos(\theta)$-angular dependence of the mean quantum oscillation frequency $\overline{F}$ for a 2D Fermi surface.}
\label{fig:RawData}
\end{figure*}

The calculated and ARPES-measured Fermi surfaces of PdRhO$_2$ are compared in Fig. \ref{fig:ARPES-FS}. The ARPES measurements yield a Luttinger count of 0.94(4) electrons per formula unit. Similar to ARPES measurements of other metallic delafossites \cite{Kushwaha15,Sobota13}, this is slightly smaller than the half-filled band expected from electron counting, which is likely due to a small shift of the Fermi level arising from a polar surface charge. 
Nonetheless, apart from some small distortions related to details of the experiment \cite{SOM}, the measured Fermi surface is in good agreement with the projection of that calculated from density-functional theory on to the two-dimensional Brillouin zone, if they are scaled to the same total area. The calculations indicate a highly two-dimensional Fermi surface, entirely consistent with sharp spectral line widths observed in the ARPES which rule out significant $k_\mathrm{z}$ dispersion. These therefore show that the interplane dispersion in PdRhO$_2$ is extremely small; the de Haas-van Alphen effect is one of the few experimental probes capable of resolving the resulting $k_\mathrm{z}$ dependent features in the Fermi surface \cite{Bergemann03}.

In Figure \ref{fig:RawData} we show background subtracted magnetic torque data for a selection of magnetic field angles $\theta$ with respect to the crystallographic c-axis within the $\mathrm{Z}\Gamma\mathrm{L}$-plane. Strong quantum oscillations are visible for all magnetic field angles (see Fig. \ref{fig:RawData}a) and b)).  The $1/\cos(\theta)$ angular dependence of the two quantum oscillation frequencies (dashed line in Fig. \ref{fig:RawData}c) evidences the quasi-two-dimensional Fermi surface topography, while the beating of the envelope function is the first indication of out-of-plane dispersion. The lower and higher frequencies, labeled $F_\beta$ and $F_\alpha$, correspond to minimal and maximal extremal orbits respectively. These are also evident in the Fourier transforms of Fig. \ref{fig:RawData}c), which were taken over a magnetic field interval from 7.5 to $15\,\mathrm{T}$. For better accuracy the frequency splitting close to the Yamaji angles was derived from the beating envelopes. 

For $B\|c$, the mean quantum oscillation frequency $\overline{F}_0=(F_\alpha(0)+F_\beta(0))/2=26.25\,\mathrm{kT}$ is equivalent to a Fermi surface cross section of $A=2.505\,$\AA$^{-2}$. Considering the room-temperature lattice constants \cite{Kushwaha17} this corresponds to $50.3\,\%$ filling of the first Brillouin zone ($4.9814\,$\AA$^{-2}$) and Luttinger count of $1.006(10)$, where the error estimate is dominated by the likely effects of thermal contraction. Thus, in agreement with \textit{ab-initio} band structure calculations, the electronic structure of PdRhO$_2$ is described by a single half-filled band with 1.00 charge carriers per formula unit.

The effective cyclotron mass, Dingle temperature and mean free path were determined for fields close to the $c$-axis.  Details of the analysis are given in \cite{SOM}. The key results are the masses $m_\alpha=1.43(5)\,m_0$ and $m_\beta=1.63(5)\,m_0$ and the mean free path is $225(30)\,\mathrm{nm}$.

\begin{figure}[ht]
	\centering
		\includegraphics[width=\columnwidth]{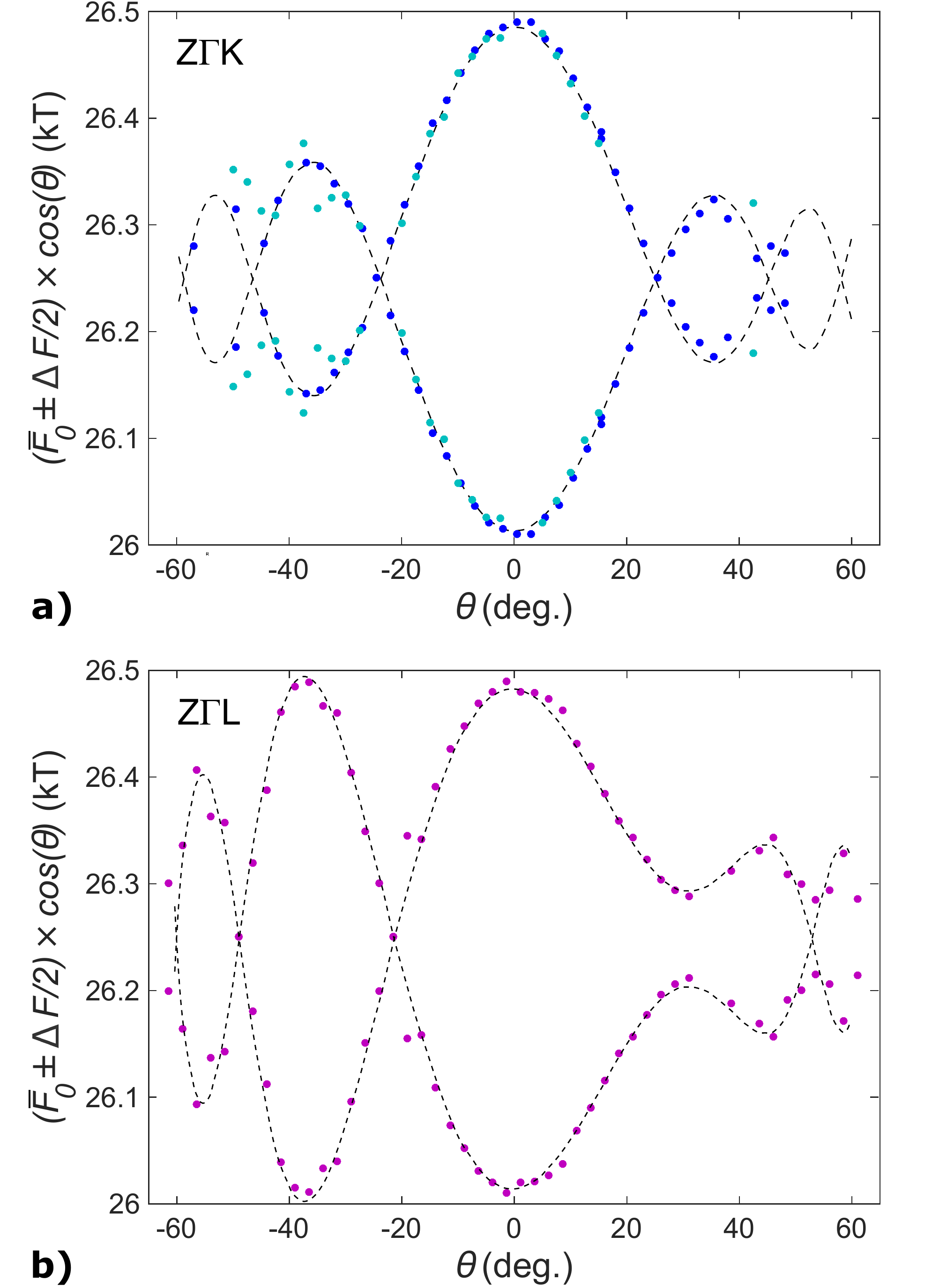}
	\caption{The graph shows the angular dependence of the quantum oscillation frequencies for magnetic field angles within the crystallographic Z$\Gamma$K-plane \textbf{(a)} and Z$\Gamma$L-plane \textbf{(b)}. Dark blue and violet symbols are data points taken on the same PdRhO$_2$ single crystal, whereas light blue symbols originate from a second sample from the same growth batch. Black dashed lines correspond to the cylindrical harmonic expansion of best fit. The associated harmonic parameters can be found in Tab. \ref{tab:SphHamExp}.}
	\label{fig:Overview}
\end{figure}

In order to analyze the Fermi surface topography further, we now turn to the angular dependence of the observed frequency splitting. The quantum oscillation frequencies for magnetic fields within the crystallographic Z$\Gamma$K and Z$\Gamma$L-planes corrected by $\cos(\theta)$ are shown in Fig. \ref{fig:Overview}. Only the frequency splitting around the mean frequency $\overline{F}$ is shown, as the angular inaccuracy of our rotator leads to sizable frequency offsets especially at larger angles. For the raw data and detailed analysis of the angular uncertainty see \cite{SOM}. 

The Fermi surface warping, i.e. azimuthal and height dependence of $k_F$, can be parametrised in cylindrical harmonics:
\begin{eqnarray}
k_\mathrm{F}=\sum_{\mu,\nu\geq0}{k_{\mu,\nu}\cos(\nu\kappa)\cos(\mu\phi)},
\label{eqn:warping}
\end{eqnarray}
where $\kappa=c^*k_z$ is the reduced $z$-coordinate and $\phi$ the azimuthal angle \cite{Bergemann00}. Note that $c^*=6.034\,\text{\AA}$ is the interlayer spacing, which is a third of the c-axis lattice constant. Due to the hexagonal lattice symmetry and $R\overline{3}m(D^5_{3d})$ space group, $k_{\mu,\nu}$ are limited to $(\mu,\nu) \in \{(0,0);(0,1);(0,2);(0,3);(3,1);(6,0);(12,0)\}$ and higher order terms. By fitting to the frequencies shown in Fig. \ref{fig:Overview}, as described in detail in \cite{SOM}, we are able to determine $k_{0,0}$ and all relevant $k_{\mu\nu}$ with $\nu\geq1$. The in-plane parameters $k_{6,0}$ and $k_{12,0}$ were obtained from the Fermi surface shape of Fig. \ref{fig:ARPES-FS} \cite{SOM}. The respective parameters and Fermi surface topology are summarized in Tab. \ref{tab:SphHamExp}. 

\begin{table}[htb]
\caption{Experimentally determined cylindrical harmonic expansion parameters of PdRhO$_2$:}
\centering
\begin{tabular}{c c c}
\multicolumn{3}{c}{\includegraphics[width=0.8\columnwidth]{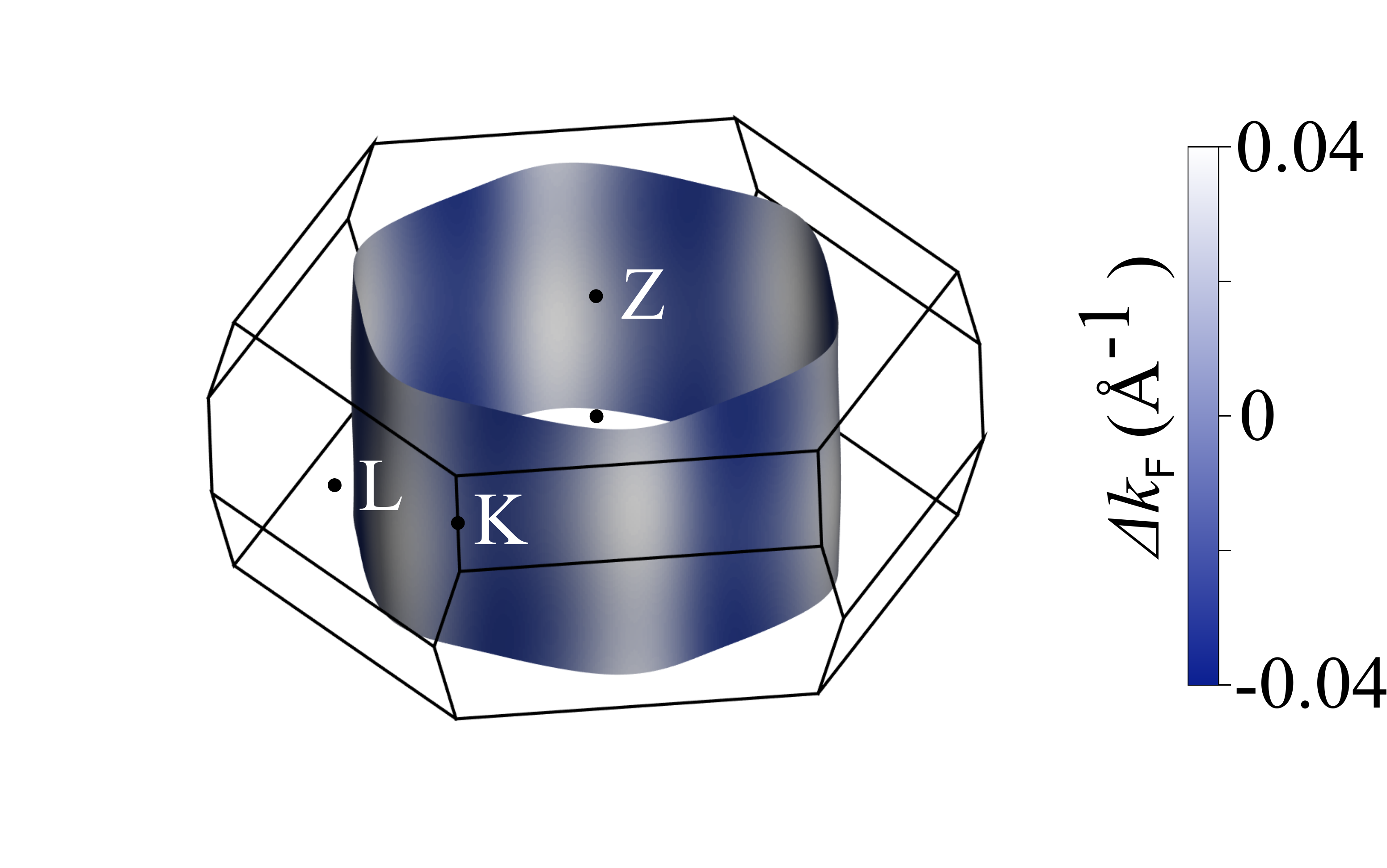}}\\
\multicolumn{3}{c}{Cylindrical Harmonic Expansion Parameters}\\

\includegraphics[width=0.25\columnwidth]{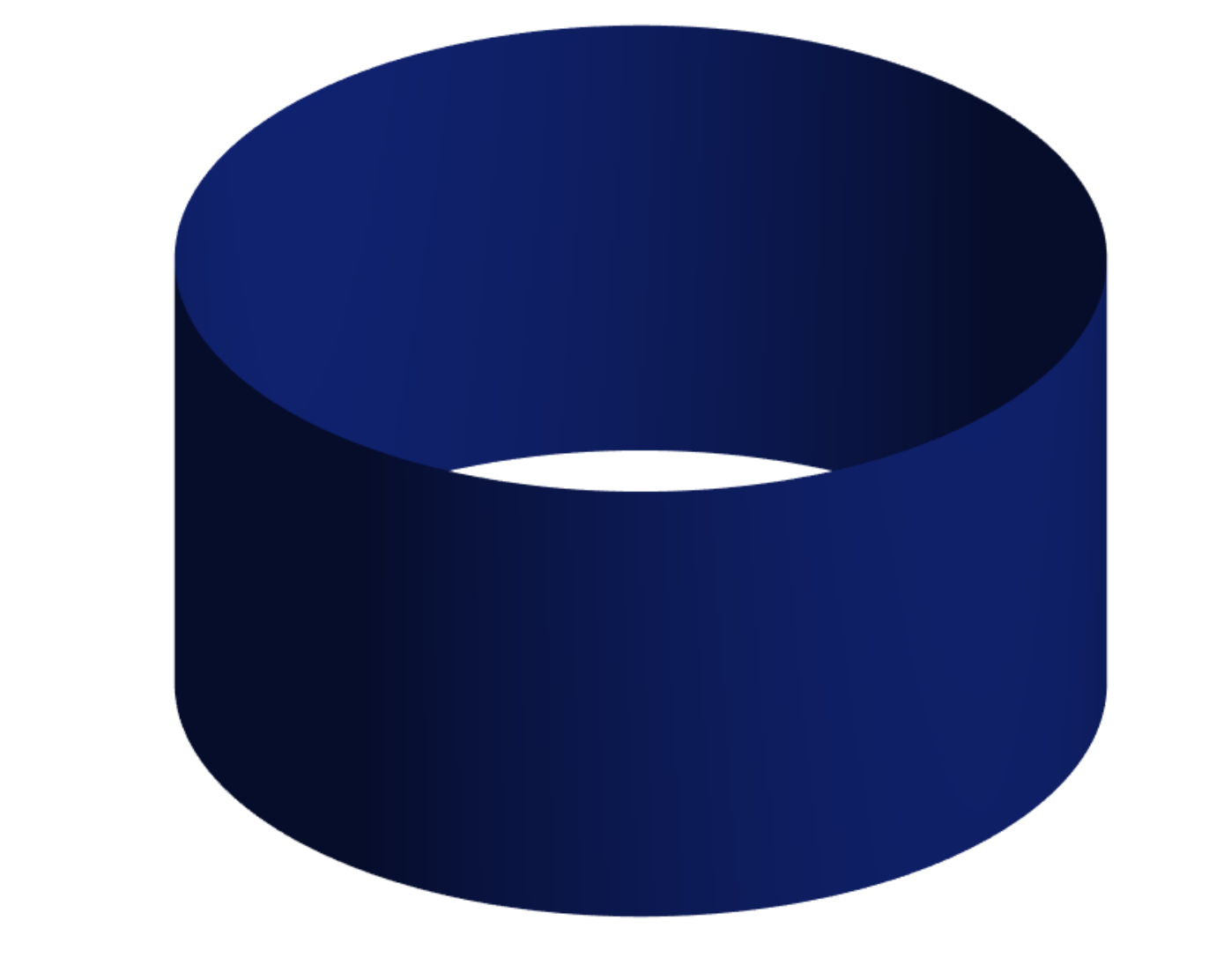} & \includegraphics[width=0.25\columnwidth]{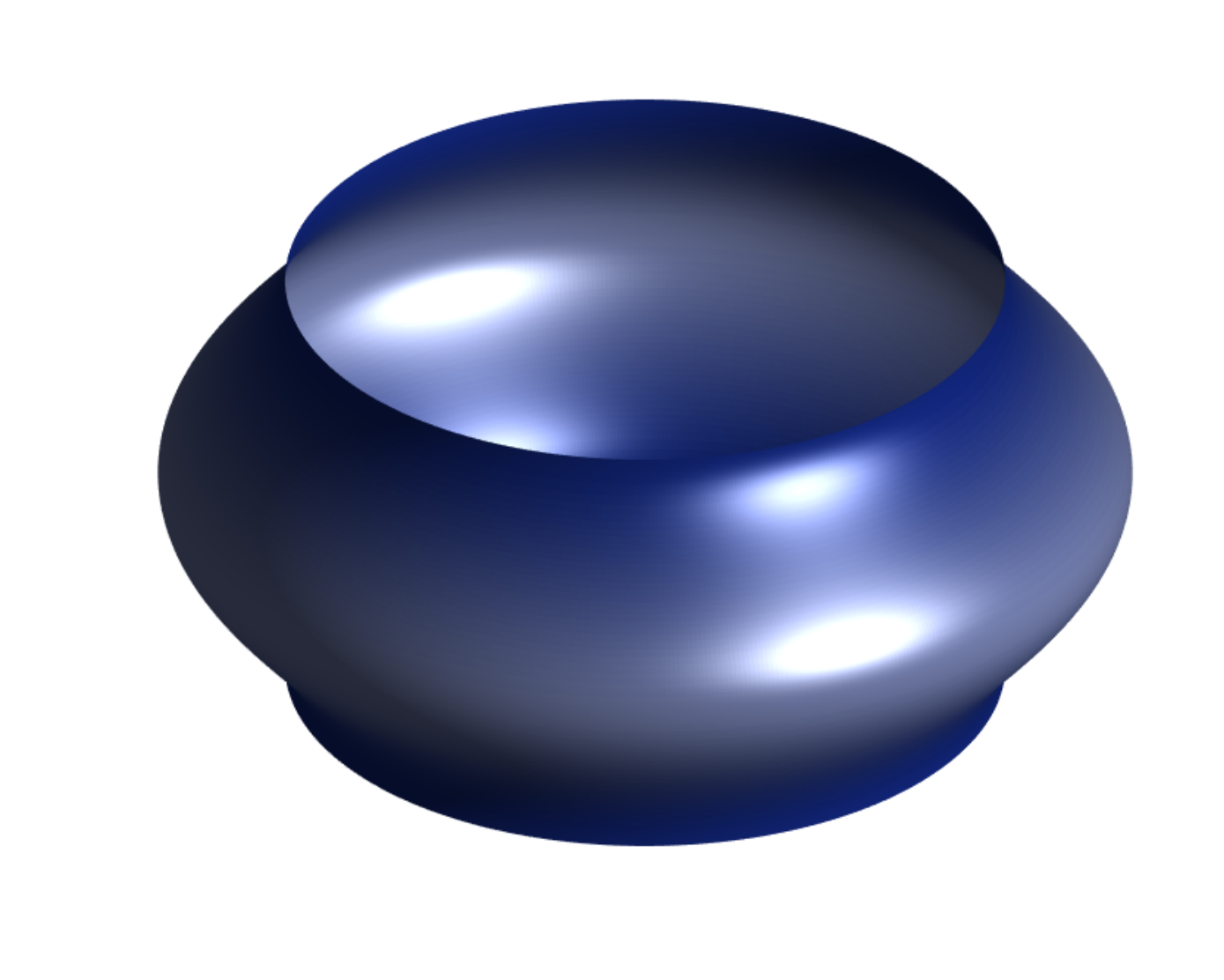} &\includegraphics[width=0.25\columnwidth]{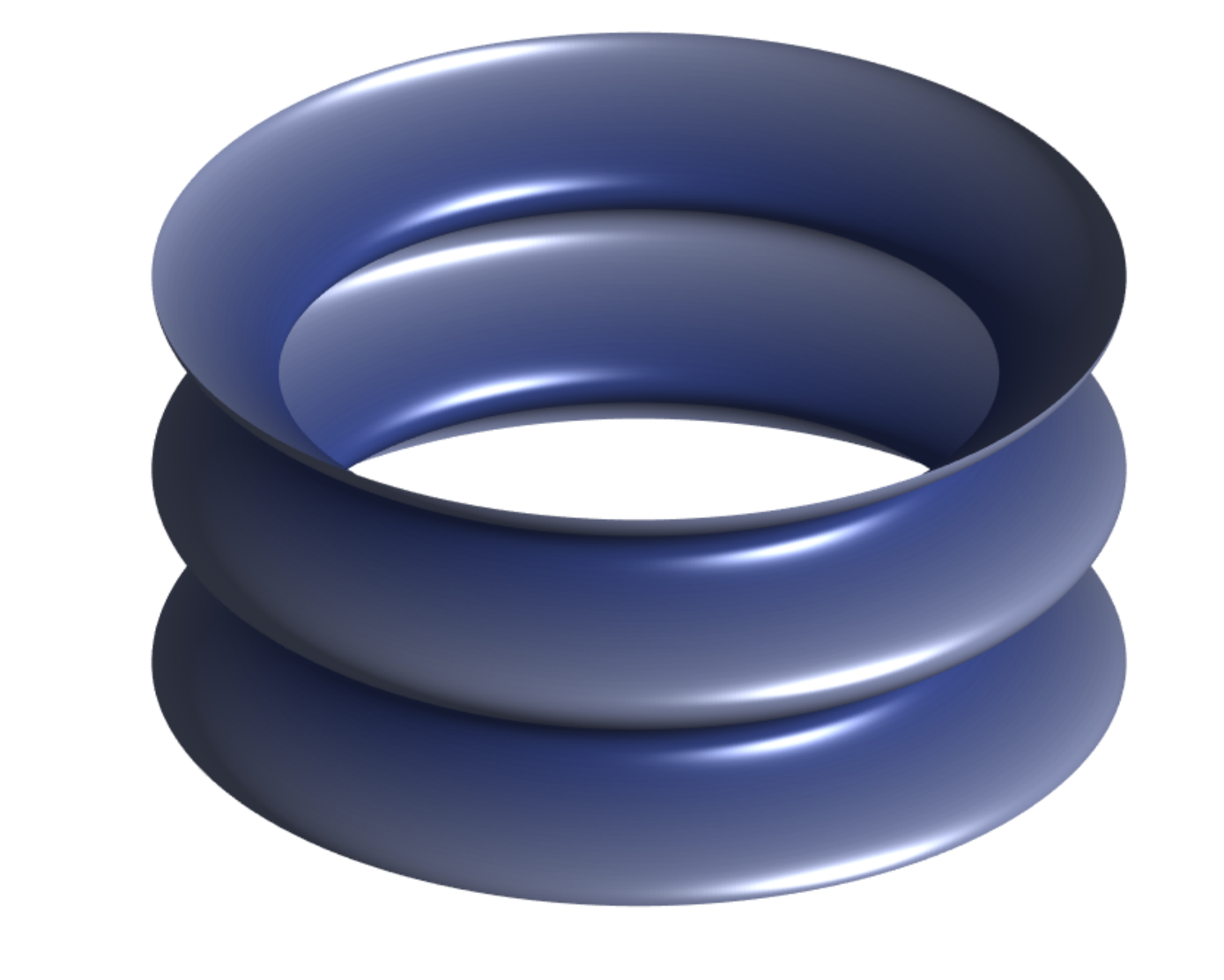} \\

$k_{0,0}$ & $k_{0,1}$ & $k_{0,2}$ \\
 0.8931(1) & 0.0040(2) & 0.0000(2) \\ 
 dHvA & dHvA & dHvA \\
\hspace{1em}\\
\includegraphics[width=0.25\columnwidth]{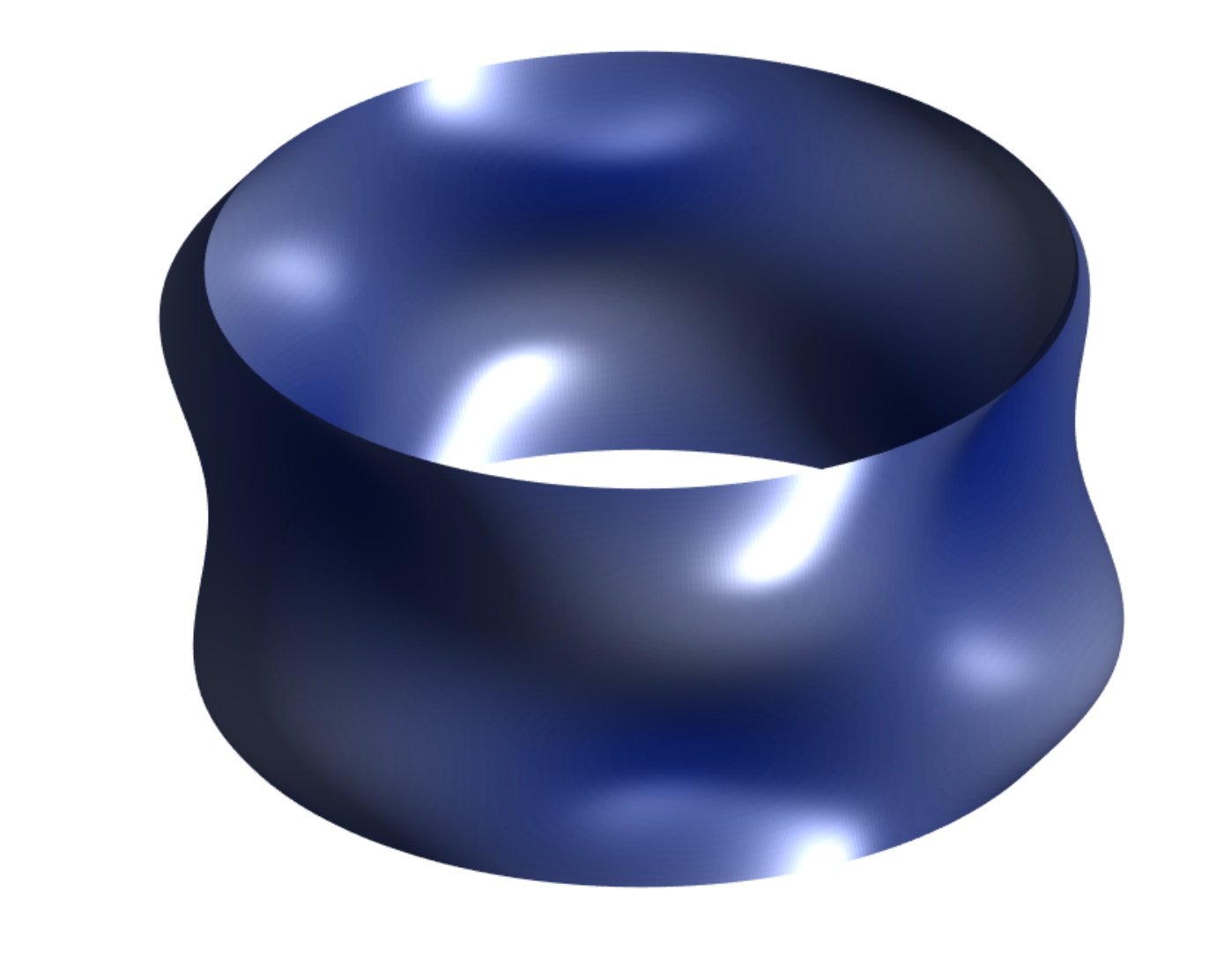} & \includegraphics[width=0.25\columnwidth]{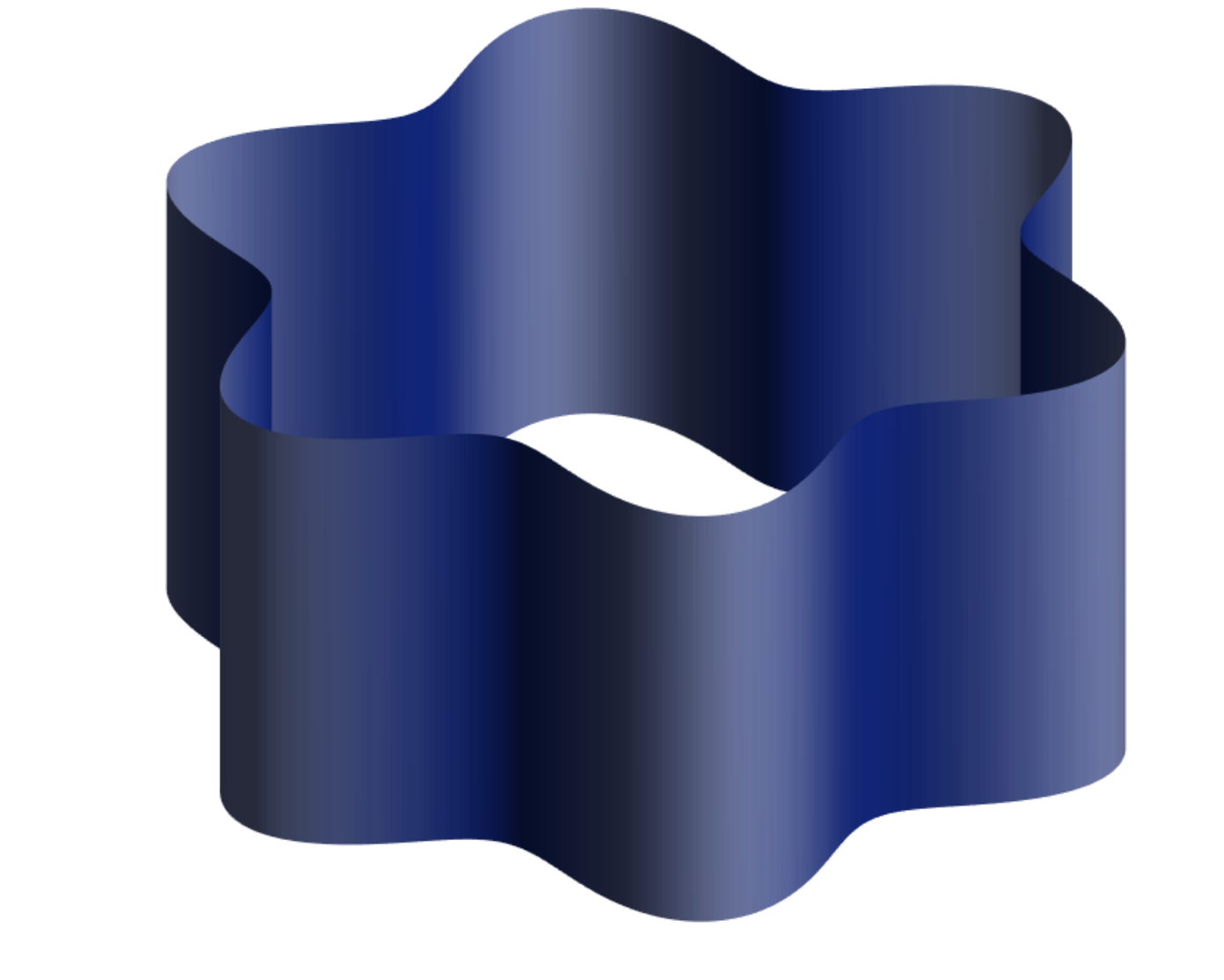} &\includegraphics[width=0.25\columnwidth]{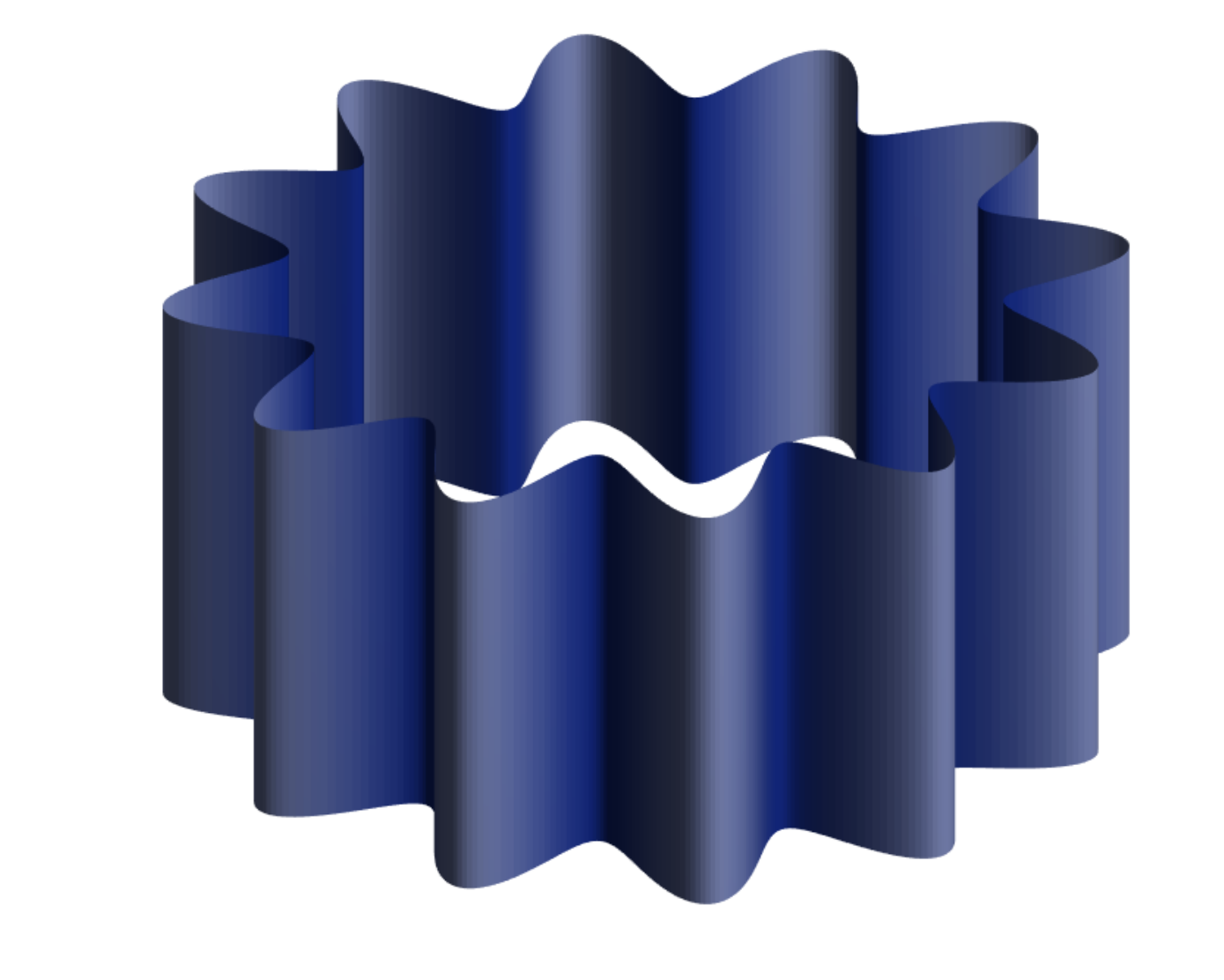} \\ 
$k_{3,1}$ & $k_{6,0}$ & $k_{12,0}$ \\
 0.0060(2) & 0.028(6) & 0.002(2) \\ 
 dHvA & ARPES \cite{Sunko17}& ARPES \cite{Sunko17} \\
\end{tabular}
\label{tab:SphHamExp}
\end{table}

Knowledge of the warping parameters of PdRhO$_2$ and a comparison with those previously deduced for its sister compound PdCoO$_2$ \cite{Hicks12} yields considerable insight into interplane hopping and coherence in the metallic delafossites. In both materials the dominant interplane terms are $k_{0,1}$, qualitatively corresponding to direct Pd-Pd hopping along the $c$-axis, and $k_{3,1}$, which results from hopping via the Co or Rh layers. In going from Co to Rh, several effects are expected to compete. Rh is larger, with more extended $4d$ orbitals, so its presence increases the in-plane $a$ and interplane $c$ lattice parameters, by approximately $7\,\%$ and $2\,\%$ respectively. This lattice expansion would be expected to lead to less effective $c$-axis Pd-Pd hopping, consistent with the observation that $k_{0,1}$ is a factor of 2.7 smaller in PdRhO$_2$ than in PdCoO$_2$. For hopping via the Co/Rh layer the situation is more subtle. If correlations in that layer are ignored, an LDA calculation predicts a much larger $k_{3,1}$ term in PdCoO$_2$ than is actually observed. However, if some account is taken of that correlation by assuming a realistic on-site repulsion energy $U$ of several eV \cite{Hicks12,SOM}, the hybridization with the conduction band is strongly suppressed, reducing the calculated value to close to the experimental one of $k_{3,1}=0.001$. Qualitatively, the lattice parameter expansion caused by moving from Co to Rh, which naively would be expected to reduce $k_{3,1}$, is more than offset by the reduction in $U$ for the $4d$ states of Rh and an increase in Pd-Rh overlap. The result is a slightly larger value of $k_{3,1}=0.006$. 

Overall, the Fermi surface of PdRhO$_2$ is extremely anisotropic, and the most two-dimensional of any metallic delafossite. Under the assumption of a single scattering time $\tau$, the $k_{\mu,\nu}$ harmonics can be used to estimate the resistive anisotropy. For a single band metallic delafossite with an assumed circular Fermi surface (the hexagonal cross-section of Fig. \ref{fig:ARPES-FS} only alters this estimate by a few per cent) the relevant expression is 
\begin{eqnarray}
\frac{\rho_\mathrm{ab}}{\rho_\mathrm{c}}=\frac{d^2}{2} \sum_{\nu,\mu>0}{\nu^2 k_{\mu,\nu}^2 \left(1+\delta_{\mu0} \right)}
\end{eqnarray}
where $d$ is the interlayer spacing and $\delta$ the Kronecker delta function. Since $k_{0,1}$ contributes more strongly to this sum than $k_{3,1}$, PdRhO$_2$ is predicted to have a larger anisotropy $(\approx1300)$ than PdCoO$_2$. Preliminary transport data \cite{Kushwaha17} are consistent with this prediction, though a more careful transport study with a range of sample sizes is desirable. The larger size of Rh also affects the in-plane Pd-Pd overlaps and reduces $k_{6,0}$, $k_{12,0}$ and the Fermi velocity $v_\mathrm{F}$. Using $k_{0,0}$ and the measured masses leads to a Brillouin zone averaged Fermi velocity $\overline{v_\mathrm{F}}=\hbar k_{0,0}/m^* = 6.8\times10^5\,\mathrm{ms}^{-1}$. This is smaller than that of PdCoO$_2$ by approximately $10\,\%$, consistent with the $a$ lattice parameter being $7\,\%$ larger in PdRhO$_2$.

Although it is possible to qualitatively account for the trends of the warping harmonics and Fermi velocity on going from PdCoO$_2$ to PdRhO$_2$, the resolution of the data that we have presented provides a considerable opportunity to refine the quality of electronic structure calculations. Despite the lower correlation energies for $4d$ Rh and Pd than for $3d$ transition metals, correlation still plays an important role in determining the details of the observed Fermi surface, and in tuning the degree of interlayer hopping. Knowing the experimental warpings at $0.1\,\%$ resolution presents a considerable challenge to "\textit{ab initio} plus correlation" theoretical approaches. It will be intriguing to see if any are capable of accounting for the values that we report for $k_{6,0}$, $k_{12,0}$, $k_{0,1}$, $k_{0,2}$, $k_{3,1}$ and $v_\mathrm{F}$. Although this seems a difficult task, PdRhO$_2$ will be an ideal material on which to benchmark the progress of the field. Preliminary attempts to add a single $U$ on the Rh site were not successful in matching all the parameters simultaneously; refinement at the level of individual Wannier functions is likely to be necessary.

A further property of note is the extremely high overall anisotropy of the measured Fermi surface. If sufficiently high anisotropies can be obtained in very clean materials like the metallic delafossites, it is possible that at high magnetic fields a limit could be reached in which all electrons are restricted to a single Landau level of very high index. Hence the physics of singly occupied Landau levels, long thought to be restricted to low density electron gases, might be observable at full metallic electron densities. Although the total bandwidth along $k_\mathrm{z}$ in as-grown PdRhO$_2$ is very small, it is still $40\,\mathrm{meV}$, implying that a field of nearly $500\,\mathrm{T}$ would be required to reach this limit. However, this observation provides motivation to try to produce a still more anisotropic material, perhaps using uniaxial pressure in PdRhO$_2$ or by growing crystals of the next compound in the series, PdIrO$_2$. This latter material is also of considerable interest as a candidate triangular lattice superconductor.  

In summary, we have successfully established the Fermi surface topography of the metallic delafossite PdRhO$_2$, using a combination of angle-resolved photoemission spectroscopy and high resolution torque magnetometry studies of the de Haas-van Alphen effect. Our results establish it as a benchmark material for the study of high purity quasi-two dimensional metals, and for the development of high precision electronic structure calculations.

\section{Acknowledgments}

The authors would like to thank the Diamond Light Source for access to Beamline I05 via Proposal No.~SI14927 as well as L. Bawden, T.K. Kim, and M. Hoesch for their technical support. In addition we would like to acknowledge the financial support from the European Research Council (through the QUESTDO project), the Engineering and Physical Sciences Research Council UK (Grant No. EP/I031014/1 and EP/L015110/1), the Royal Society and the Max-Planck Society.


%

\clearpage

\section*{Supplementary Online Material - Quasi Two-Dimensional Fermi Surface Topography of the Delafossite PdRhO$_2$}

\subsection{1. Sample Preparation and Characterization}

\subsubsection{1.1 Samples}

\noindent Figure~\ref{fig:Samples} shows the two samples whose quantum oscillations were studied in this article. Both samples were mounted on PRC400 piezo-electric micro-cantilevers \cite{Hitachi} with Apiezon N-grease. Using grease allows us to reorient the sample on the cantilever, whilst it forms a solid bond at low temperatures. 

\subsubsection{1.2 Laue Diffraction}

\begin{figure}[b]
	\centering
			\includegraphics[width=0.9\textwidth]{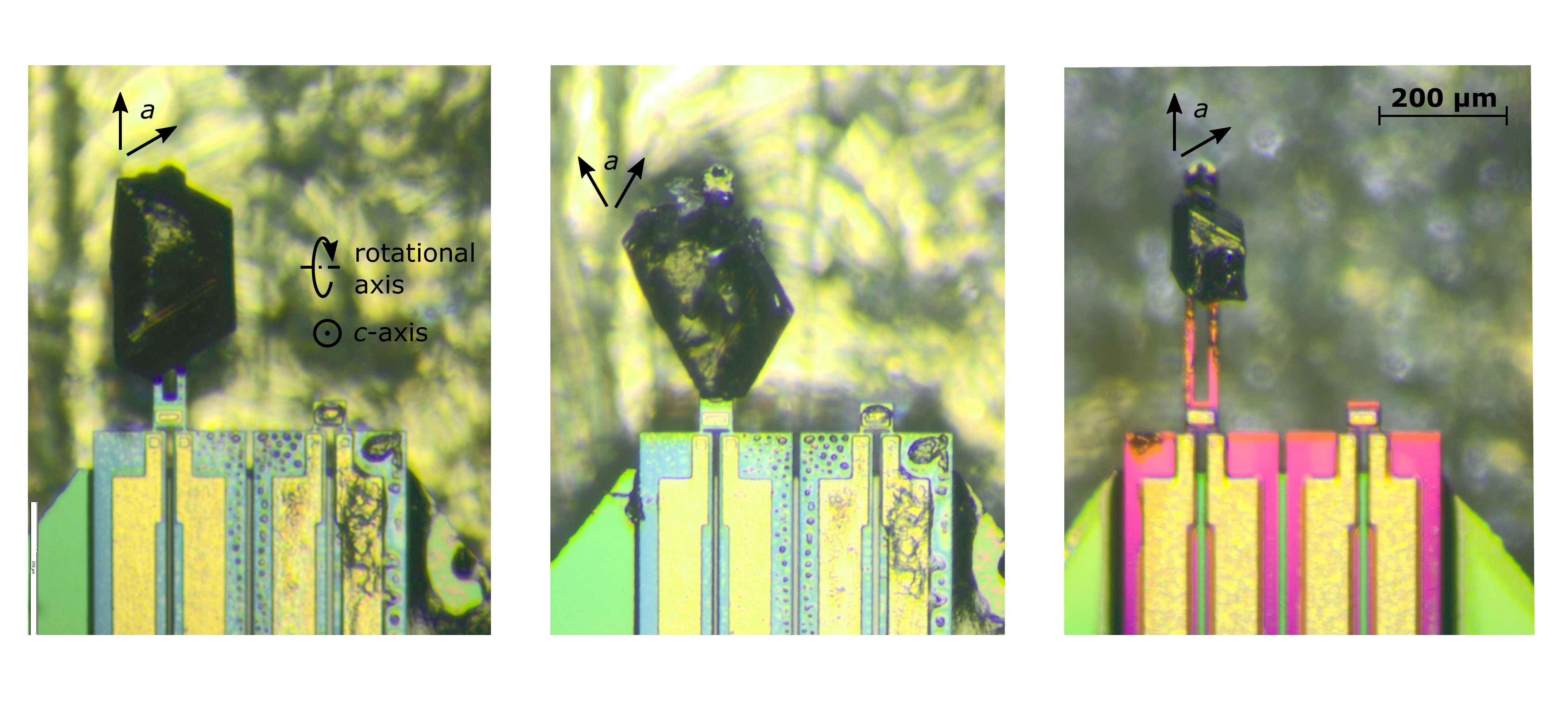}
	\caption{Photos of two PdRhO$_2$ samples mounted on $400\,\mu\mathrm{m}$ piezo-resistive micro-cantilevers for magnetic torque measurements. \textbf{a)} and \textbf{b)} show large sample \#1 mounted for magnetic fields within the $\mathrm{Z}\Gamma\mathrm{K}$ and $\mathrm{Z}\Gamma\mathrm{L}$-plane respectively \textbf{c)} shows a second, much smaller sample \#2 from the same growth batch mounted for measurements within the $\mathrm{Z}\Gamma\mathrm{K}$-plane. $a$ and $c$ refer to a hexagonal representation of the rhombohedral structure.}
	\label{fig:Samples}
\end{figure}

\begin{figure}[tbp]
	\centering
		\includegraphics[width=0.90\columnwidth]{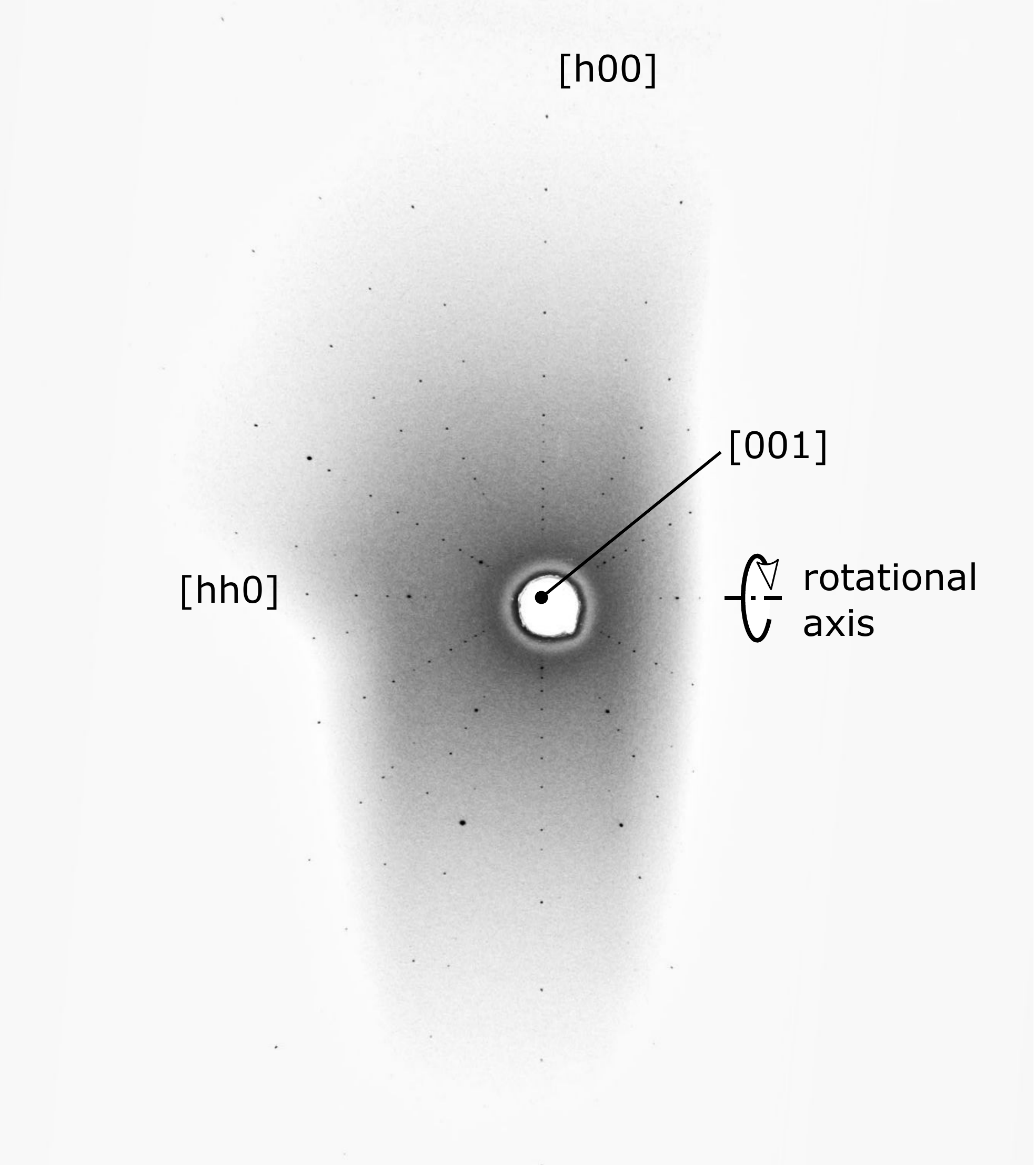}
	\caption{Laue diffractogram of the large PdRhO$_2$ single crystal in the orientation shown in Fig.~\ref{fig:Samples}b). The rotation axis is as indicated by the dash-dotted line.}
	\label{fig:LaueSLarge}
\end{figure}

\noindent Laue x-ray diffractograms (Fig.~\ref{fig:LaueSLarge}) of the large PdRhO$_2$ single crystal ($200\times 300 \times 50\,\mu\mathrm{m}$) in the orientiation shown in Fig.~\ref{fig:Samples}b) were taken to confirm its single crystallinity and orientation. For this diffraction measurement the sample was still mounted on the micro-cantilever and silver sample holder used in the rotation study of the dHvA oscillations. A collimator of $0.5\,\mathrm{mm}$ diameter was used, probing the entire sample at once.

The diffractogram (Fig.~\ref{fig:LaueSLarge}) only shows higher order Bragg peaks of PdRhO$_2$. The [100] and [110] reflections (note that these are the hexagonal representation of the rhombohedral unit cell) are masked by the silver sample holder to the left and right of the image. However, the Bragg peaks show the single crystallinity and orientation of the sample. The rotational axes for the angular dependence of the dHvA frequencies are shown in the Laue pattern (Fig.~\ref{fig:LaueSLarge}) and sample photos (Fig.~\ref{fig:Samples}). Thus, in the orientation shown in Fig.~\ref{fig:Samples}b), the sample is rotated around the [110] direction corresponding to magnetic fields within the $\mathrm{Z}\Gamma\mathrm{L}$ plane.

\onecolumngrid

\newpage

\twocolumngrid

\subsection{2. Angle-Resolved Photo Emission Spectroscopy (ARPES)}

\noindent As stated in the main article, the in-plane warping parameters $k_{6,0}$ and $k_{12,0}$ of the PdRhO$_2$ Fermi surface were extracted from a slightly distorted Fermi surface contour.

The measured ARPES Fermi surface displayed in Fig.~1 of the main text exhibits slight distortions from the expected 3-fold rotational symmetry within the surface Brillouin zone, which we attribute to the presence of slight surface inhomogeneity, small positioning errors from the center of rotation of the sample manipulator, or possibly small residual fields or local work-function variations introducing distortions on the outgoing electron trajectories. The data shown in Fig.~1 of the main text were used to establish the quoted $k_{6,0}$ and $k_{12,0}$, but numerically correcting the distortions makes only a tiny quantitative change to the extracted values.

\begin{figure}[b]
	\centering
		\includegraphics[width=0.95\columnwidth]{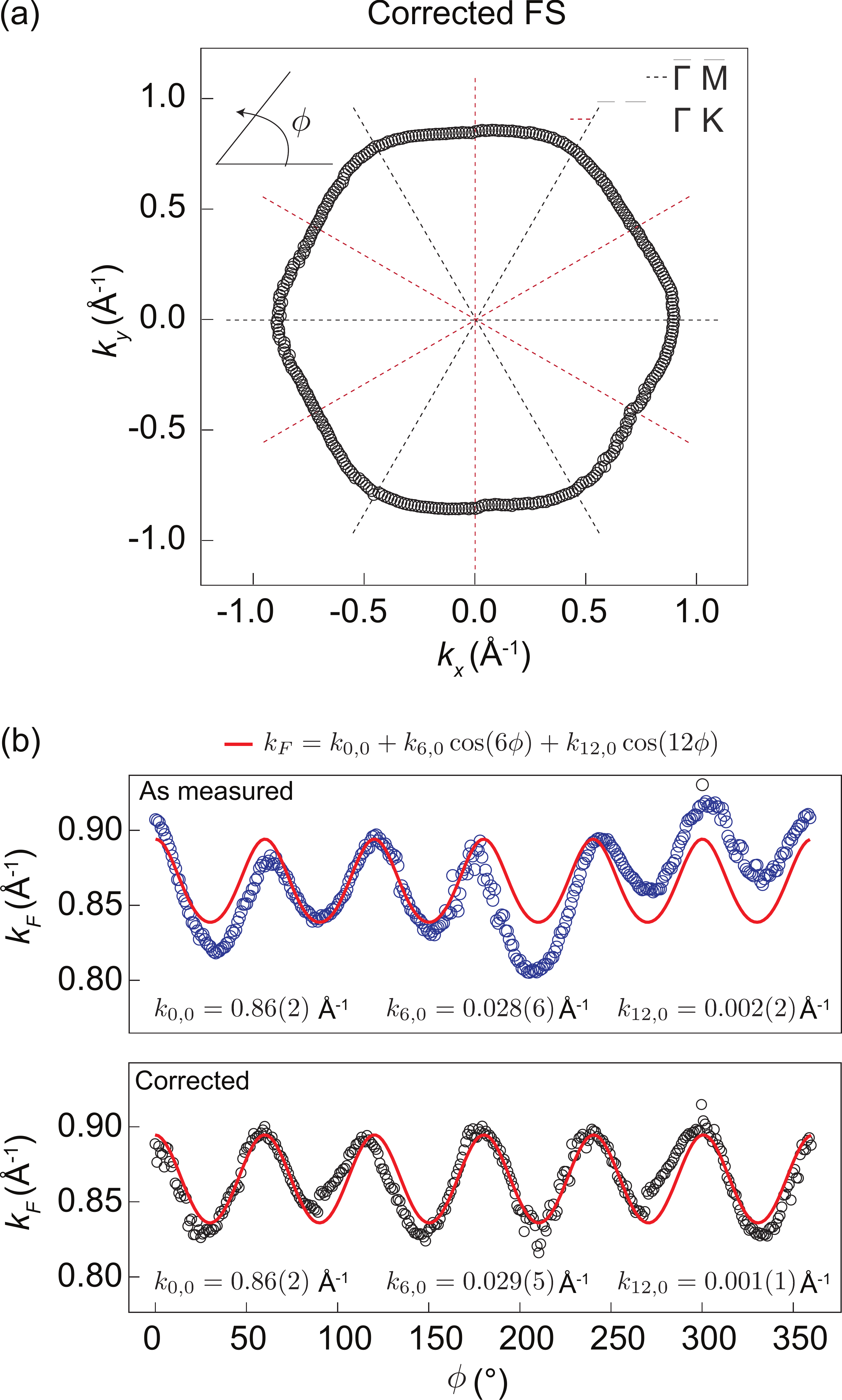}
	\caption{(a) Fermi surface of PdRhO$_2$ extracted from ARPES measurements, after correction of the angular distortions. This was achieved by fitting Fermi momenta (each averaged over a $4^\circ$ azimuthal bin width) at 24 points equally spaced around the Fermi surface, and enforcing these to be symmetric at time-reversal equivalent points of the surface Brillouin zone (i.e. $\pm{k_F}=\pm0.5(k_+-k_-)$). This yielded a set of anchor points for an angular remapping of the measured Fermi surface. The extracted Fermi momenta from angular cuts around the Fermi surface is shown (b) before and (c) after this distortion correction. Fits to the in-plane harmonic components are shown as red lines. These yield almost identical values with and without applying the distortion correction.}
	\label{fig:ARPES-Harm}
\end{figure}

Fig.~\ref{fig:ARPES-Harm} shows the Fermi surface as determined by ARPES and the corresponding azimuthal dependence of the Fermi wave vector. As can be seen the in-plane warping-parameters stay constant within the error bars when the distortion is numerically corrected.

\clearpage

\subsection{3. de Haas-van Alphen (dHvA)}

\subsubsection{3.1 Effective Mass Analysis}

\begin{figure}[b]
	\centering
		\includegraphics[width=1.00\columnwidth]{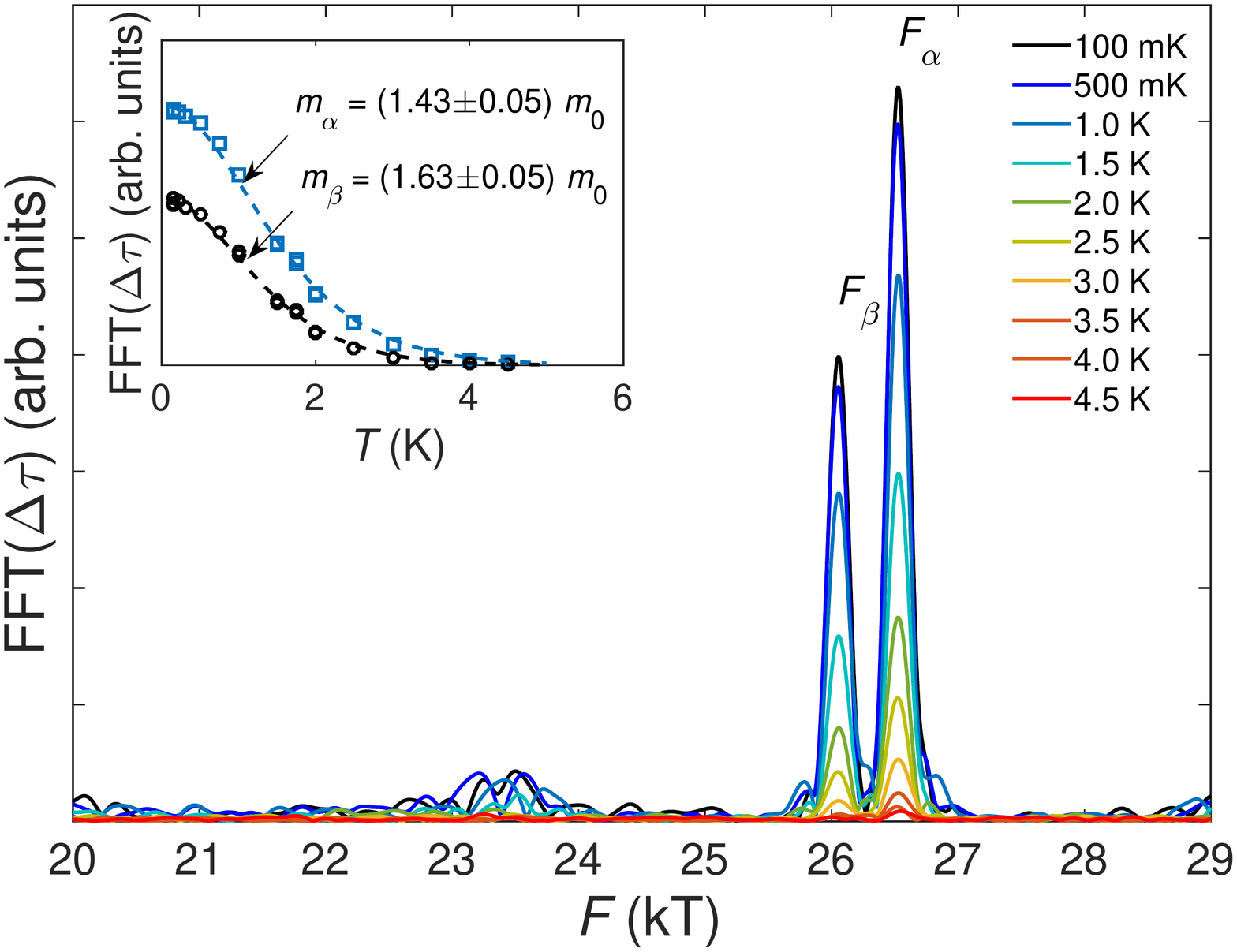}
\caption{Quantum oscillation spectra at a magnetic field angle of $5^\circ$ within the $\mathrm{Z}\Gamma\mathrm{L}$-plane for temperatures between $100\,\mathrm{mK}$ and $4.5\,\mathrm{K}$. Fourier transforms were taken in the magnetic field window of 13 to $15\,\mathrm{T}$. The inset shows the temperature dependencies of the two quantum oscillation frequencies and the according fit to the Lifshitz-Kosevich temperature reduction term \cite{Shoenberg,Lifshitz56}.}
\label{fig:TempDep}
\end{figure}
	
\noindent Cyclotron masses of both extremal orbits were determined from a temperature dependence of the quantum oscillation amplitude at a magnetic field angle of $+5^\circ$ within the $\mathrm{Z}\Gamma\mathrm{L}$-plane. Figure~\ref{fig:TempDep} shows the measured quantum oscillation spectra for temperatures between $100\,\mathrm{mK}$ and $4.5\,\mathrm{K}$. 
Due to the poor thermal conductance of the micro-cantilevers at millikelvin temperatures, sample temperatures below $200\,\mathrm{mK}$ were calculated according to \cite{Arnold17RSI} using the stabilized rotator temperature and excitation current of $10\,\mu\mathrm{A}$.
 As can be seen the two dHvA frequencies are strongly suppressed with increasing temperature. Both temperature dependencies are well described by the Lifshitz-Kosevich temperature reduction term:
\begin{eqnarray}
R_\mathrm{T}=\frac{x}{\sinh\{x\}} \hspace{1em}\mathrm{with}\hspace{1em} x=\frac{\pi^2m^*k_\mathrm{B}T}{\mu_\mathrm{B}B}
\end{eqnarray}
(see inset of Fig.~\ref{fig:TempDep} of the main text) The corresponding cyclotron masses are $m_\beta=(1.63\pm0.05)\,m_0$ ($26.06\,\mathrm{kT}$) and $m_\alpha=(1.43\pm0.05)\,m_0$ ($26.53\,\mathrm{kT}$).

\subsubsection{3.2 Dingle Analysis}

\begin{figure}[b]	
  \centering		
		\includegraphics[width=1.00\columnwidth]{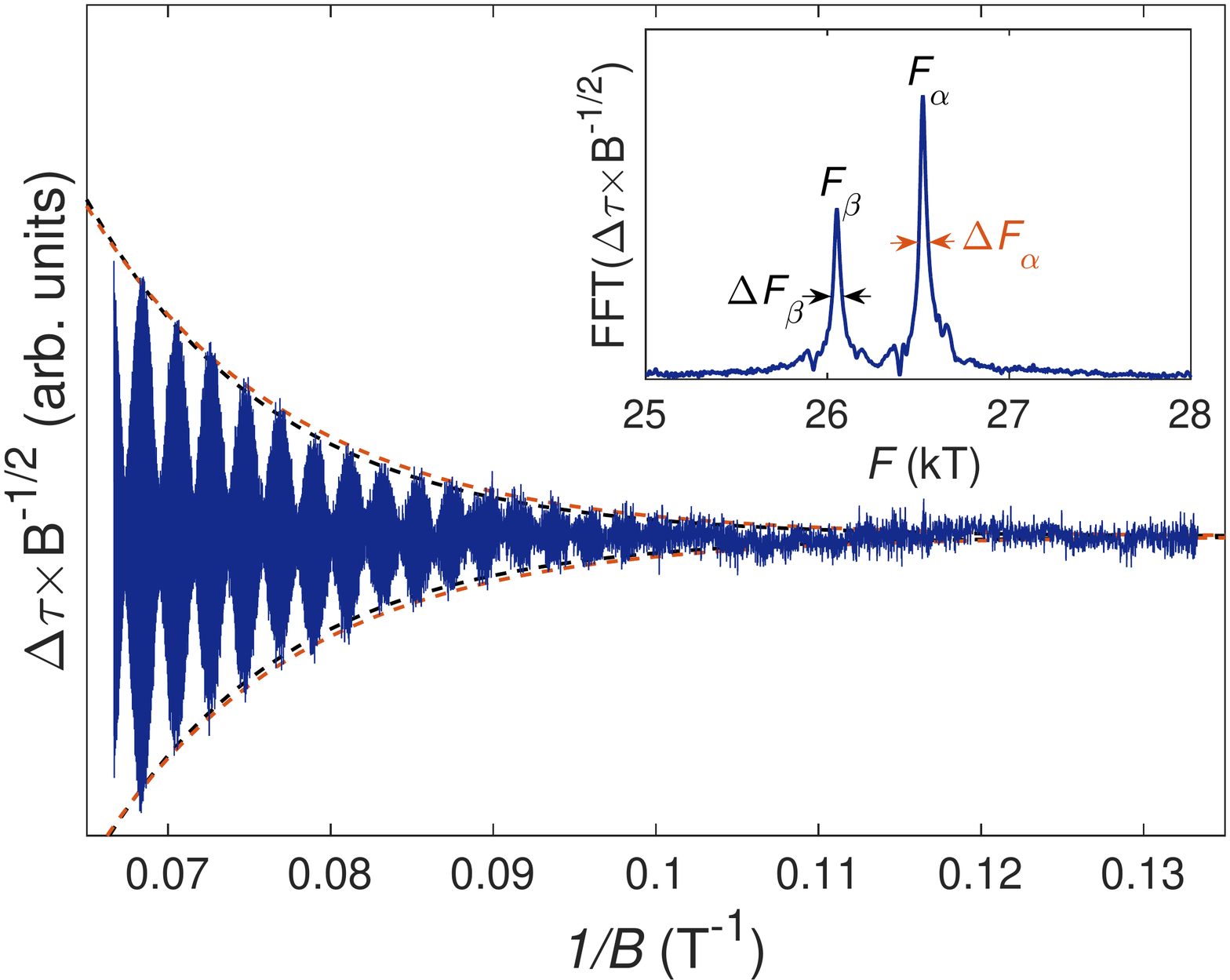}
\caption{Dingle magnetic field dependence of the quantum oscillation amplitude at $100\,\mathrm{mK}$ corrected by $\sqrt{B}$ \cite{Dingle52}. The Dingle temperature and scattering times for both orbits are inferred from the line width of the quantum oscillation spectrum shown in the inset. The dashed lines show the Dingle envelopes corresponding to the inferred line widths and effective masses.}
	\label{fig:FieldDep}
\end{figure}

\noindent Information about the mean free path and charge carrier scattering times can be drawn from the magnetic field dependence of the quantum oscillation amplitude. The magnetic field dependence is described by the Dingle term: 
\begin{eqnarray}
R_\mathrm{D} = \exp\left\{-\frac{\pi^2m^*k_\mathrm{B}T_\mathrm{D}}{\mu_\mathrm{B}B}\right\},
\end{eqnarray}
where $T_\mathrm{D}$ is the Dingle temperature, which is indirect proportional to the scattering time $\tau =\hbar/(2\pi k_\mathrm{B} T_\mathrm{D})$. As strong beating occurs in PdRhO$_2$, a direct fit of the Dingle term to the quantum oscillation envelope is subject to large errors. Thus, we pursue an alternative approach and determine the Dingle temperature from the line width of the quantum oscillation spectrum Fig.~\ref{fig:FieldDep}b). For this the torque data are corrected by $\sqrt{B}$ to account for the intrinsic magnetic field dependence of the Lifshitz-Kosevich equation. The exponentially decaying envelope transforms in a Lorenzian line shape, whose full-width-half-maximum is $\Delta F = T_D \pi m^*k_\mathrm{B}/\mu_\mathrm{B}$ \cite{Arnold17ArXiv}. From the Dingle spectrum Fig.~\ref{fig:FieldDep}, we extract Dingle temperatures of $T^\beta_\mathrm{D}=3.6\,\mathrm{K}$ and $T^\alpha_\mathrm{D}=3.8\,\mathrm{K}$ and scattering times of $\tau^\beta = 3.4\times10^{-13}\,\mathrm{s}$ and $\tau^\alpha = 3.2\times10^{-13}\,\mathrm{s}$ respectively. Taking into account the extremal Fermi surface cross sections and cyclotron masses, we obtain Fermi velocities of $v^\beta_\mathrm{F}=633\,\mathrm{km/s}$ and $v^\alpha_\mathrm{F}=725\,\mathrm{km/s}$ ($v_\mathrm{F}=\hbar k_\mathrm{F}/m^*$) and electron mean free path of $\overline{l} = \tau \times v_\mathrm{F}\approx (225\pm10)\,\mathrm{nm}$.

\subsubsection{3.3 Angular Dependence}

\begin{figure}[tb]
	\centering
		\includegraphics[width=0.95\columnwidth]{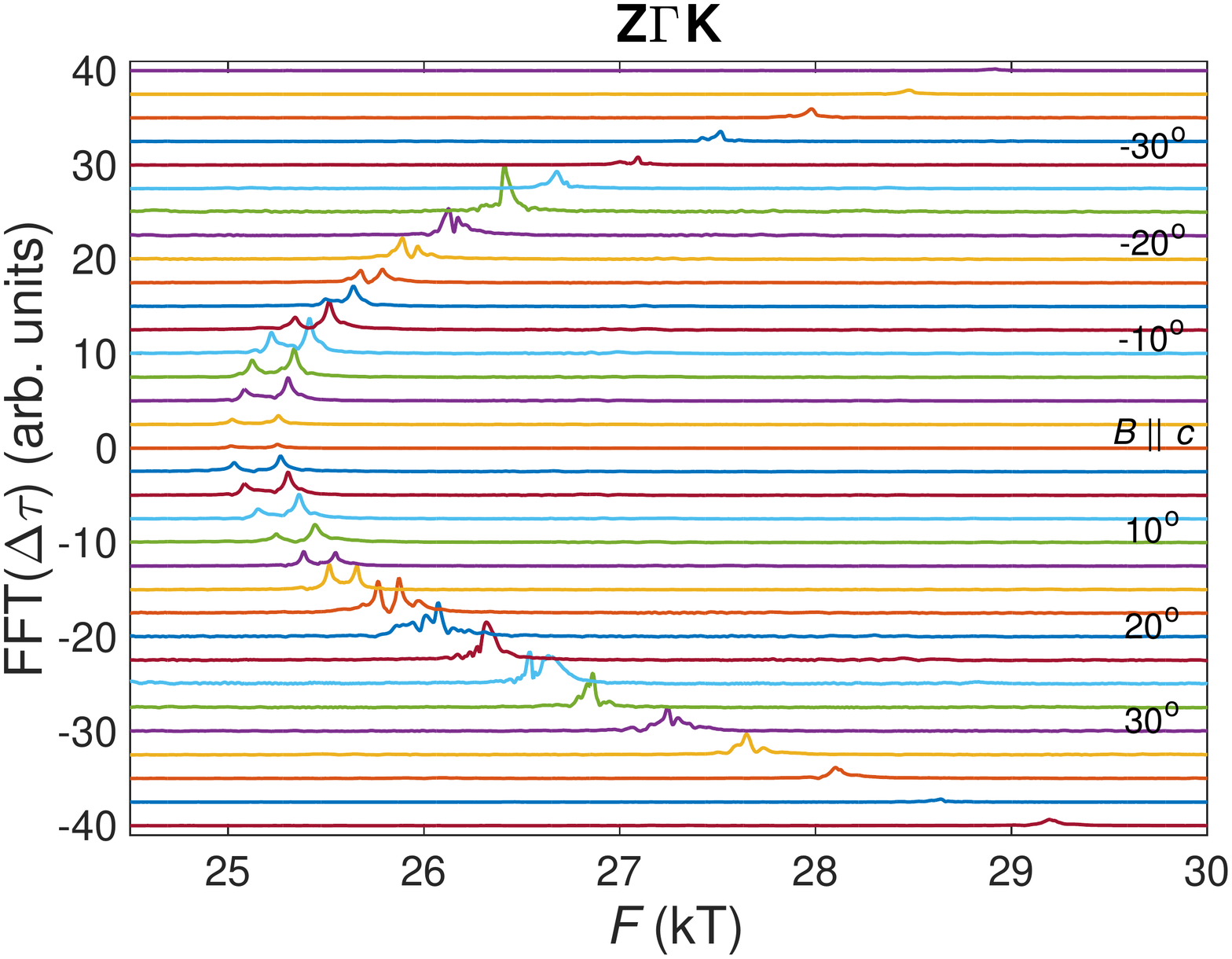}
		\centering
		\includegraphics[width=0.95\columnwidth]{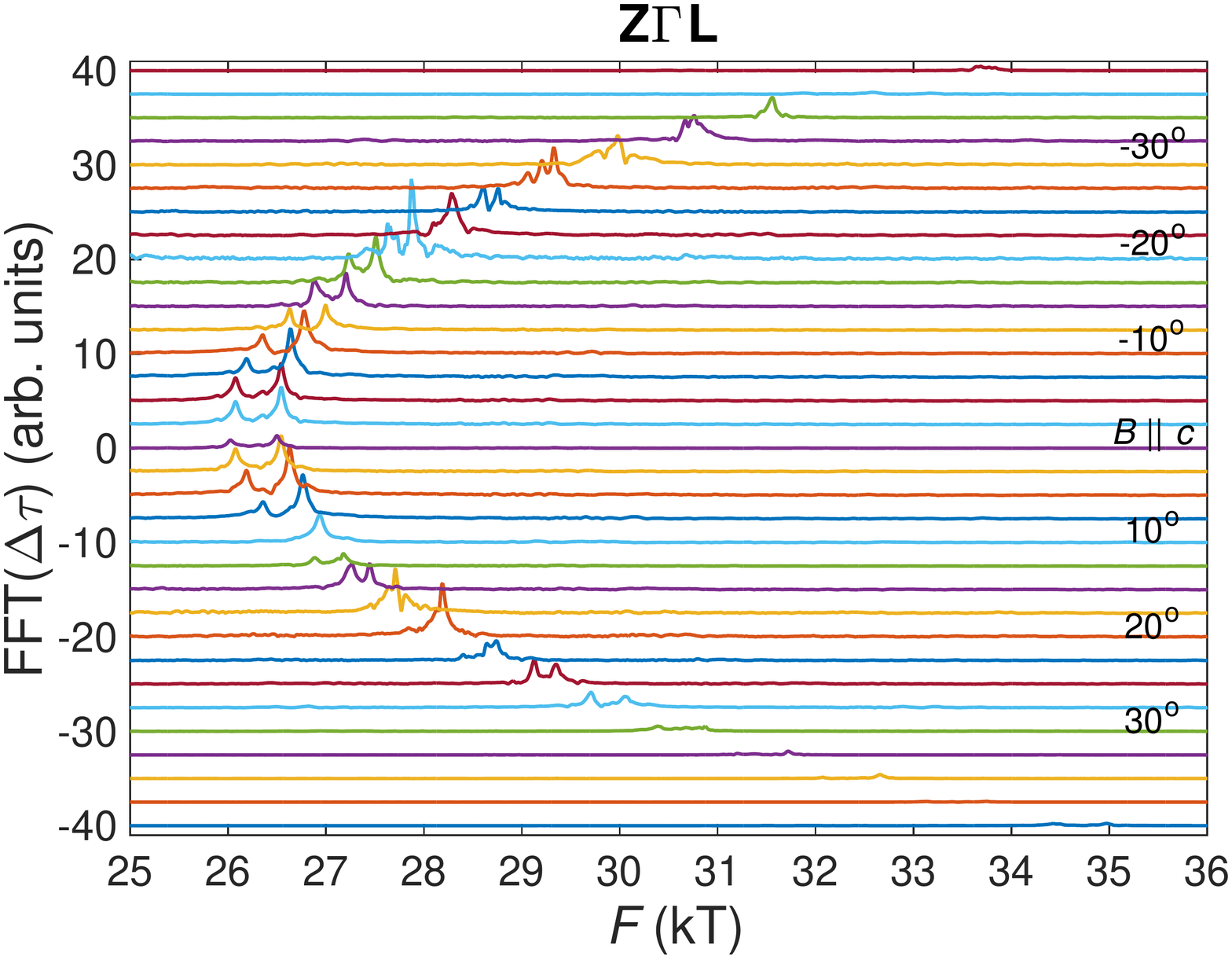}	
			\caption{Quantum oscillation spectra as determined from Fourier transforming $\Delta\tau$ over $1/B$ in the magnetic field interval of $B_\mathrm{low}=7.5\,\mathrm{T}$ to $B_\mathrm{up}=15\,\mathrm{T}$ after multiplication with $\exp(-\alpha/B)$ where $\alpha=5(B_\mathrm{up}-B_\mathrm{low})$ to suppress low field noise. Data are offset by units of their angle for clarity.}
	\label{fig:FFTvTheta_Raw}
\end{figure}

\noindent To determine the angular dependence of the quantum oscillation frequencies, magnetic torque measurements were performed between $7.5\,\mathrm{T}$ and $15\,\mathrm{T}$ in angular steps of $2.5^\circ$ for magnetic fields applied in the $\mathrm{Z}\Gamma\mathrm{K}$ and $\mathrm{Z}\Gamma\mathrm{L}$ plane respectively. After subtracting a $B^2$-background and filtering the magnetic torque data, fast Fourier transforms (FFTs) were taken as a function of $1/B$. The resulting quantum oscillation spectra are shown in Fig.~\ref{fig:FFTvTheta_Raw}.

\begin{figure}[tb]
	\centering
		\includegraphics[width=0.95\columnwidth]{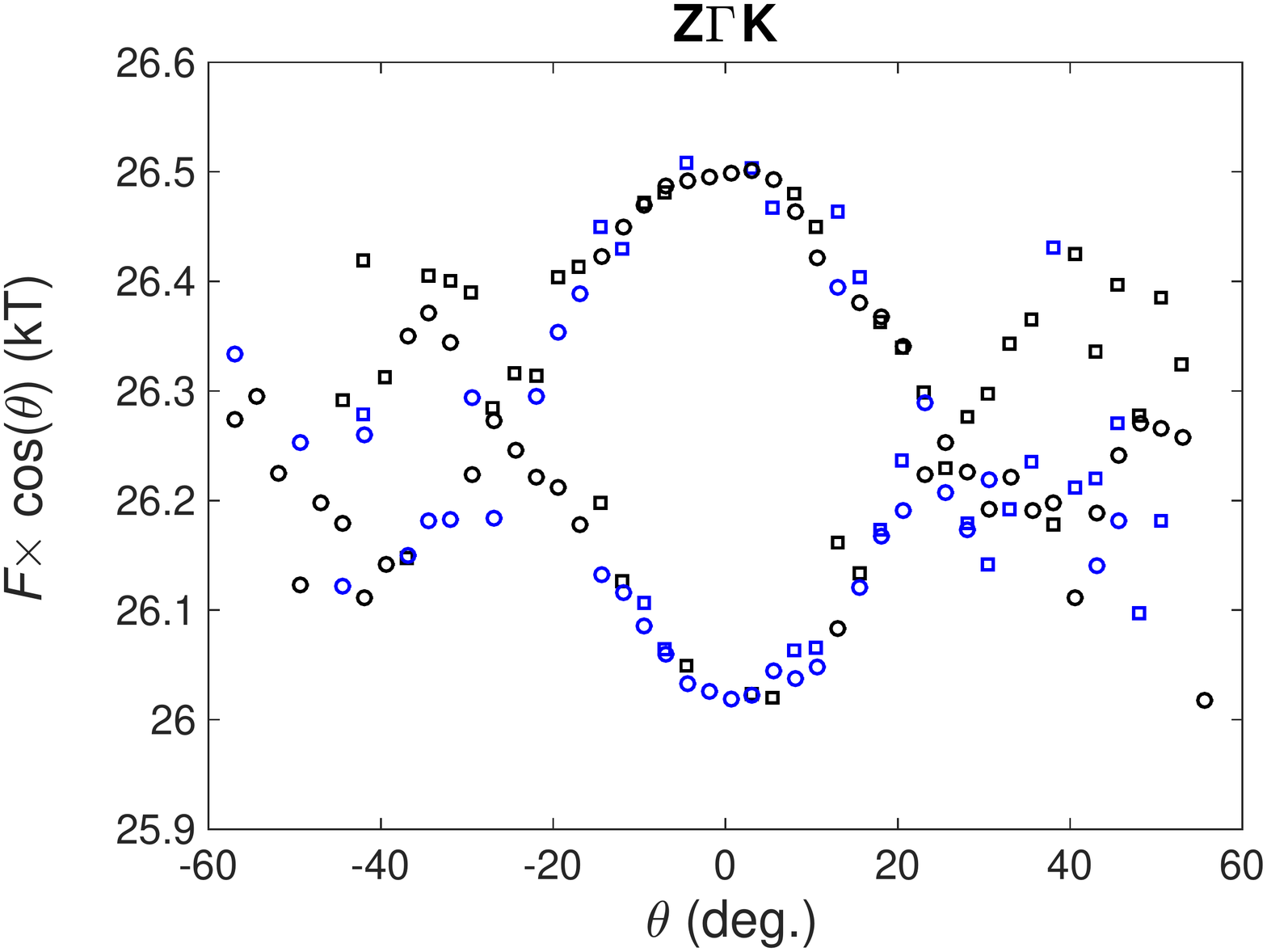}
		\centering
		\includegraphics[width=0.95\columnwidth]{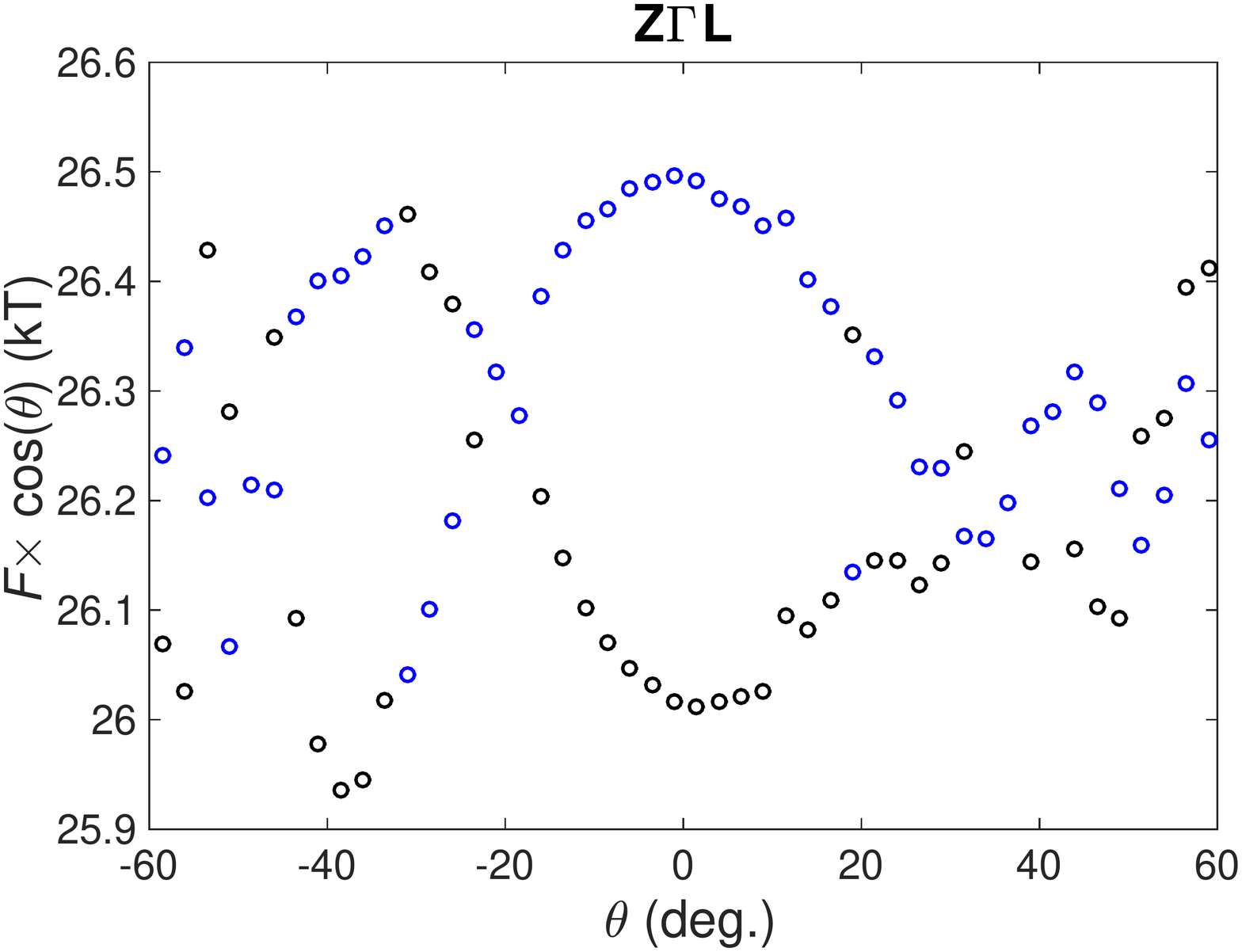}	
			\caption{Raw quantum oscillation frequencies corrected by $\cos(\theta)$. Black and blue symbols mark the strongest and weaker quantum oscillation frequencies as determined from the FFTs Fig.~\ref{fig:FFTvTheta_Raw}. Circles and Squares correspond to the larger sample \#1 and smaller sample \#2 respectively.}
	\label{fig:FCosThetavTheta_Raw}
\end{figure}

Quantum oscillation frequencies were determined from the peak positions of the FFTs and given in Fig.~\ref{fig:FCosThetavTheta_Raw}. As shown in the main article, their main angular dependance $F(\theta) = F_0 / \cos(\theta)$ is given by the quasi two dimensional shape of the Fermi surface. Here, $\theta$ is the polar angle formed by the crystal $c$ axis and the direction of the magnetic field. As we are interested in variations of the almost perfectly cylindrical shape, we correct for this angular dependence when determining the warping parameters. 

\subsubsection{3.4 Angular uncertainty}

\noindent The mounting procedure for the sample, micro-cantilever and sample holder to the rotator induces an angular uncertainty of up to $\pm2.5^\circ$ of the sample with respect to the magnetic field. Parallel alignment to the c-axis, i.e. $\theta_0$, was determined by the most symmetric $1/\cos\{\theta\}$ angular dependence. Fig.~\ref{fig:ThetaDistorsion} depicts how a small variation of the $\theta_0$ moves the quantum oscillation frequencies away from the cylindrical behavior especially for large angles.

\begin{figure}[tb]
	\centering
		\includegraphics[width=0.95\columnwidth]{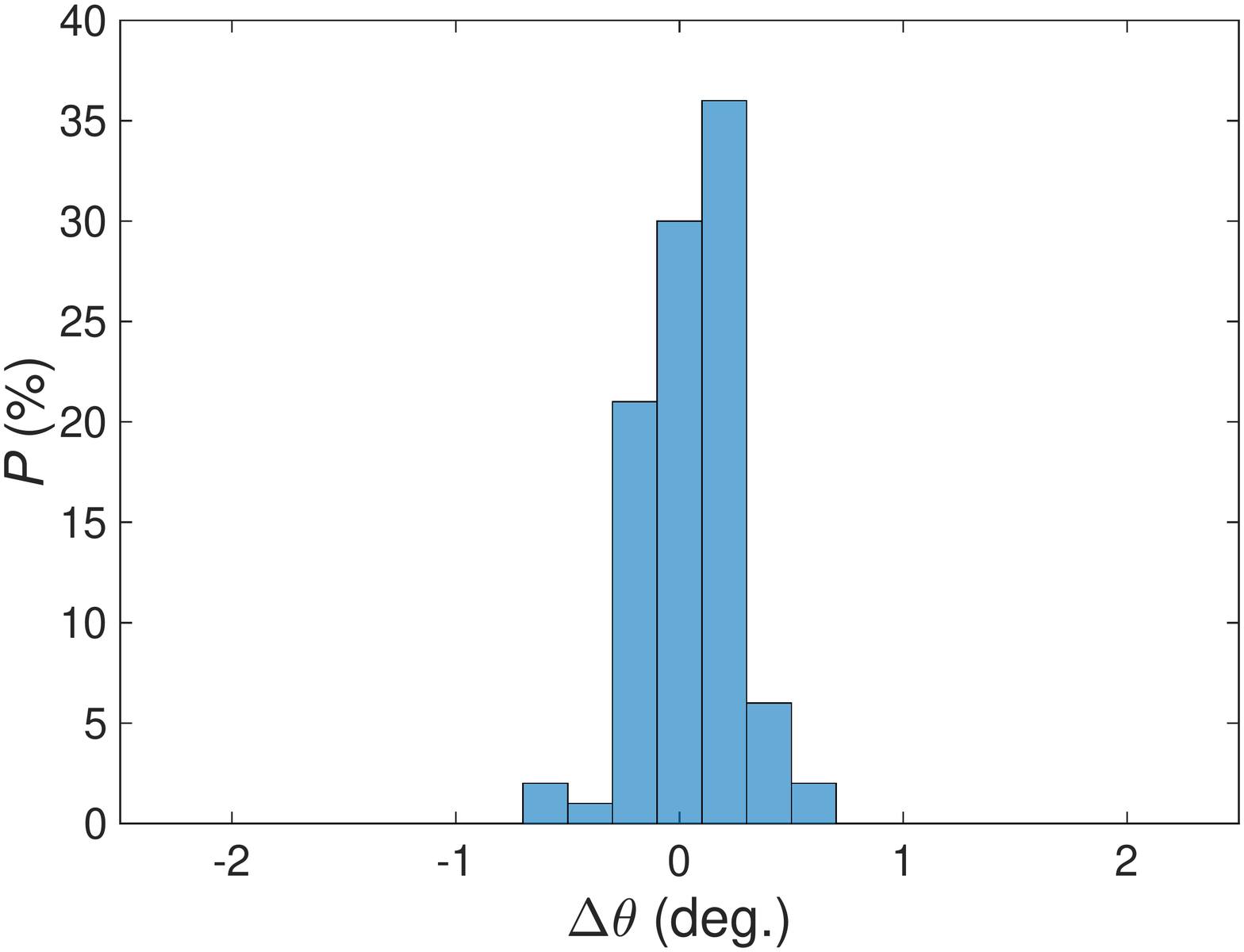}
	\caption{Histogram of the rotational error of the Swedish rotator as calculated from the discrepancy between the experimental and theoretical angular dependence of the quantum oscillation frequencies.}
	\label{fig:AngularHistogram}
\end{figure}

\begin{figure}[tb]
	\centering
		\includegraphics[width=1.0\columnwidth]{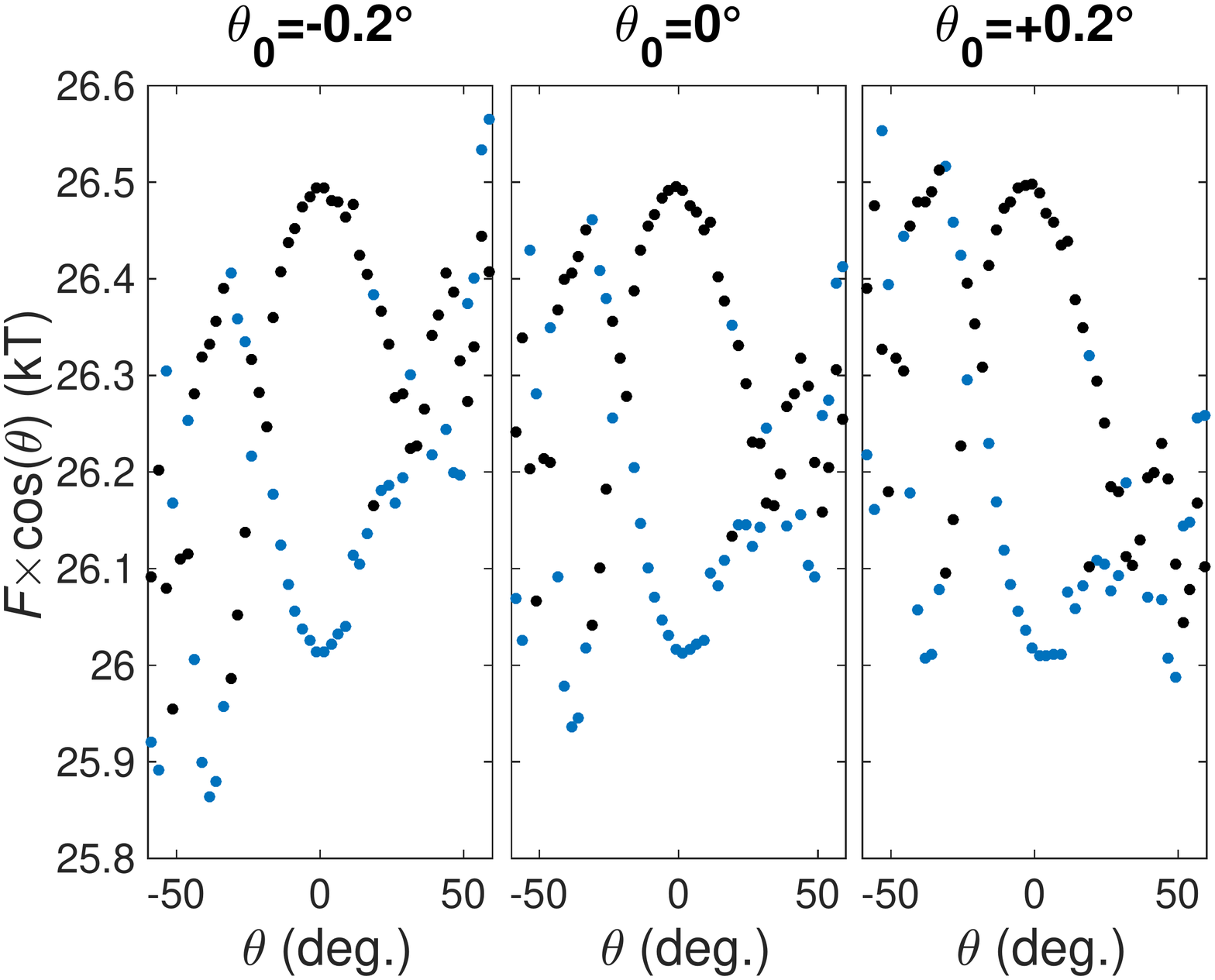}
	\caption{The graphs show how the angular dependence of the quantum oscillations within the $\mathrm{Z}\Gamma\mathrm{L}$ plane is distorted by choosing a slightly different angle for $\theta=0$.}
	\label{fig:ThetaDistorsion}
\end{figure}

Additionally to that systematic angular shift, we will discuss the angular uncertainty of our Swedish rotator as possible source of the scattering of the mean quantum oscillation frequencies. Figures~\ref{fig:FCosThetavTheta_Raw}a) and b) show the raw quantum oscillation frequencies corrected by $\cos\{\theta\}$. As can be seen the mean of the individual frequency pairs scatter around a common value of approximately $26.25\,\mathrm{kT}$. In order to quantify possible origins of this scatter, we calculate the actual angle of each frequency pair from its mean frequency and the theoretical $\overline{F}(0)/\cos\{\theta\}$-angular dependence assuming $\overline{F}(0)=26.25\,\mathrm{kT}$. A histogram of the discrepancy between the nominal and actual angle is shown in Figure~\ref{fig:AngularHistogram}.

We find that the angular discrepancy follows a standard distribution with a standard deviation of approximately $0.2^\circ$, which is in agreement with the technical specifications of our Swedish rotator. 

Scatter, originating from an angular uncertainty of the order of $0.2^\circ$ results in an error of the $\cos(\theta)$-scaling factor of approximately $0.2\,\%$ at $\theta = 30^\circ$ and up to $0.6\,\%$ at $\theta = 60^\circ$. This effect can be seen in Fig.~\ref{fig:FCosThetavTheta_Raw}, where the scatter of the mean frequency is most severe at large angles. The quoted error of up to $0.6\,\%$ induces a shift of the mean frequency of $\leq150\,\mathrm{T}$. The induced change of the frequency splitting, however, is only $\leq3\,\mathrm{T}$, which is far beneath our experimental frequency resolution. Thus the change of the frequency splitting is negligible and in Fig.~3 of the main article, the frequency splitting $\Delta F \times \cos\{\theta\}$ is given as measured whereas the mean frequency is corrected to the expected frequency $F_0$ at this angle. This introduces an artificial symmetrization of the frequency splitting, by which some information about the angular dependence of the average frequency and therefore the component $k_{0,2}$ of the harmonic expansion is lost(see below).
At larger angles, only the strongest quantum oscillation frequency is visible due to a poorer signal to noise ratio. Hence there is no information about the frequency splitting for these angles in Fig.~3 of the main article.

\subsubsection{3.5 Cylindrical harmonics expansion}


\noindent The angular dependence of the quantum oscillation frequencies due to the warping described in Eqn. 1 of the main text can be described by a Bessel function expansion of the extremal cross section $A$:
\begin{eqnarray}
A(\kappa) = \frac{\pi k^2_{00}}{\cos(\theta)}+\frac{2\pi k_{0,0}}{\cos(\theta)}&\sum_{\mu,\nu}&k_{\mu,\nu}J_\mu(\nu\kappa_\mathrm{F}\tan{\theta})\nonumber\\
&\times&\cos(\nu\kappa)\cos(\mu\phi),
\end{eqnarray}
where $\mu$ and $\nu$ are the axial and azimuthal indices of the cylindrical harmonics. The corresponding azimuthal and polar angles are $\phi$ and $\theta$. $\kappa_{F}=ck_{0,0}/3$ is the reduced planar Fermi wave vector and $J_\mu$ are the Bessel functions \cite{Bergemann00}. These cross sections are used to calculate the oscillatory part of the magnetization, i.e. the de Haas-van Alphen effect:
\begin{eqnarray}
\tilde{M} =\int^{2\pi}_0\sin\left(\frac{\hbar A(\kappa)}{eB}\right)d\kappa.
\end{eqnarray}
Fourier transforming the oscillatory magnetization in $1/B$ leads to a theoretical angular dependence of the quantum oscillation frequencies depending on $k_{\mu,\nu}$. 
 
$k_{0,0}$ is uniquely determined by the mean quantum oscillation frequency $k_{0,0}=\sqrt{2e\overline{F}_0/\hbar}$ for $B\|c$. By tuning the harmonic parameters $k_{0,1}$, $k_{0,2}$ and $k_{3,1}$, we achieved a good fit to the experimental angular dependence (dashed lines in Fig. 3 of the main text). Here, $k_{0,1}$ is mostly responsible for the frequency splitting around the $c$-axis, whereas $k_{3,1}$ determines the asymmetry between positive and negative field angles within the $\mathrm{Z}\Gamma\mathrm{L}$-plane (Fig. 3b of the main text). The parameter $k_{0,2}$ results in an asymmetry of the angular dependence of the upper versus the lower frequency branch. By a comparison of the raw data in Fig.~\ref{fig:FCosThetavTheta_Raw} with the simulation, we estimate that $k_{0,2} < 0.0002$.

Note that we had to allow for a $\approx 2^{\circ}$ azimuthal misalignment of $\mathrm{Z}\Gamma\mathrm{K}$ rotation plane to account for the observed asymmetry between positive and negative polar angles in Fig. 3a of the main text). Otherwise (for perfect alignment) the quantum oscillation frequencies within the $\mathrm{Z}\Gamma\mathrm{K}$ plane are independent of $k_{3,1}$ and symmetric about $\theta=0$.

\subsubsection{3.6 Torque interaction}

\begin{figure}[tb]
	\centering
		\includegraphics[width=0.95\columnwidth]{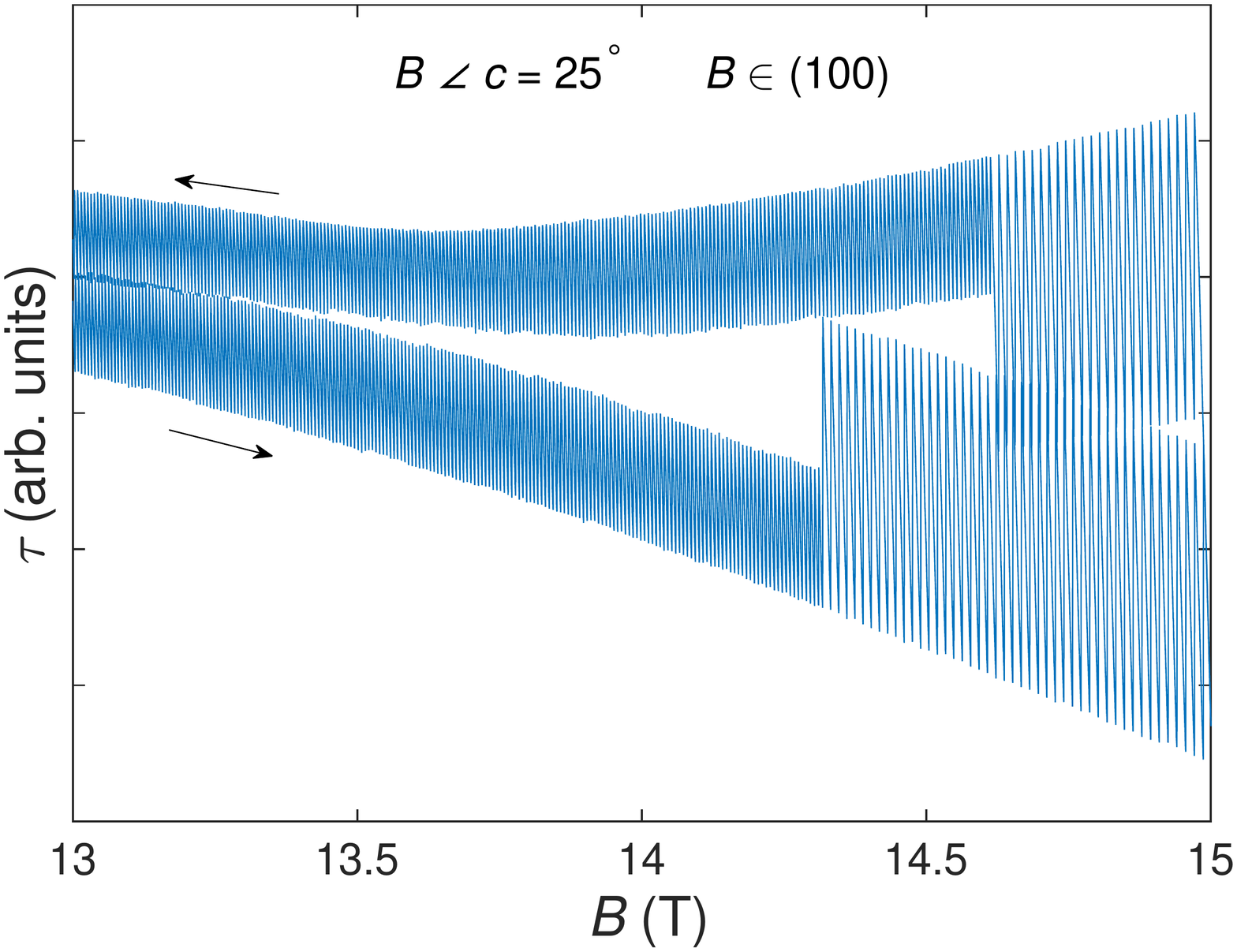}
	\caption{Magnetic torque interaction induced by a comparably large magnetic moment and deflection of the micro-cantilever.}
	\label{fig:TorqueInteraction}
\end{figure}

\noindent Close to the Yamaji angles ($\pm25^\circ$ for $B\in \mathrm{Z}\Gamma\mathrm{K}$), we found a sudden halving of the quantum oscillation frequency and doubling of the quantum oscillation amplitude for our large sample (see Fig. \ref{fig:TorqueInteraction}). The observed critical fields are highly hysteretic and only weakly depend on the magnetic field sweep rate. However they diverge quickly away from the Yamaji angle.

A closer study of this feature revealed that it is caused by magnetic torque interaction \cite{Shoenberg}. Here a large oscillatory magnetization, such as induced by the de Haas-van Alphen oscillations in our samples, causes a non-negligible deflection of the magnetic torque lever. This deflection  leads to a reduction of the effective applied magnetic field in quasi-two-dimensional materials and consequential extension of the quantum oscillation period. Due to the simultaneous crossing of multiple Landau levels through the Fermi edge in these systems, the resulting quantum oscillations are highly non-sinusoidal and appear at fractions of the original frequency.

In our case, we observe saw-tooth like quantum oscillations at highest fields and close to the Yamaji angle in the bigger sample but sinusoidal quantum oscillations in the smaller sample. Where the former shows a halving of the quantum oscillation frequency and doubling of the amplitude and the latter does not. This corroborates the torque interaction, which is larger in the larger sample, as the origin of the frequency change and excludes an intrinsic origin for its observation.

\subsubsection{3.7 Spin Zeros and $g$-factor}

\begin{figure}[tb]
	\centering
		\includegraphics[width=0.95\columnwidth]{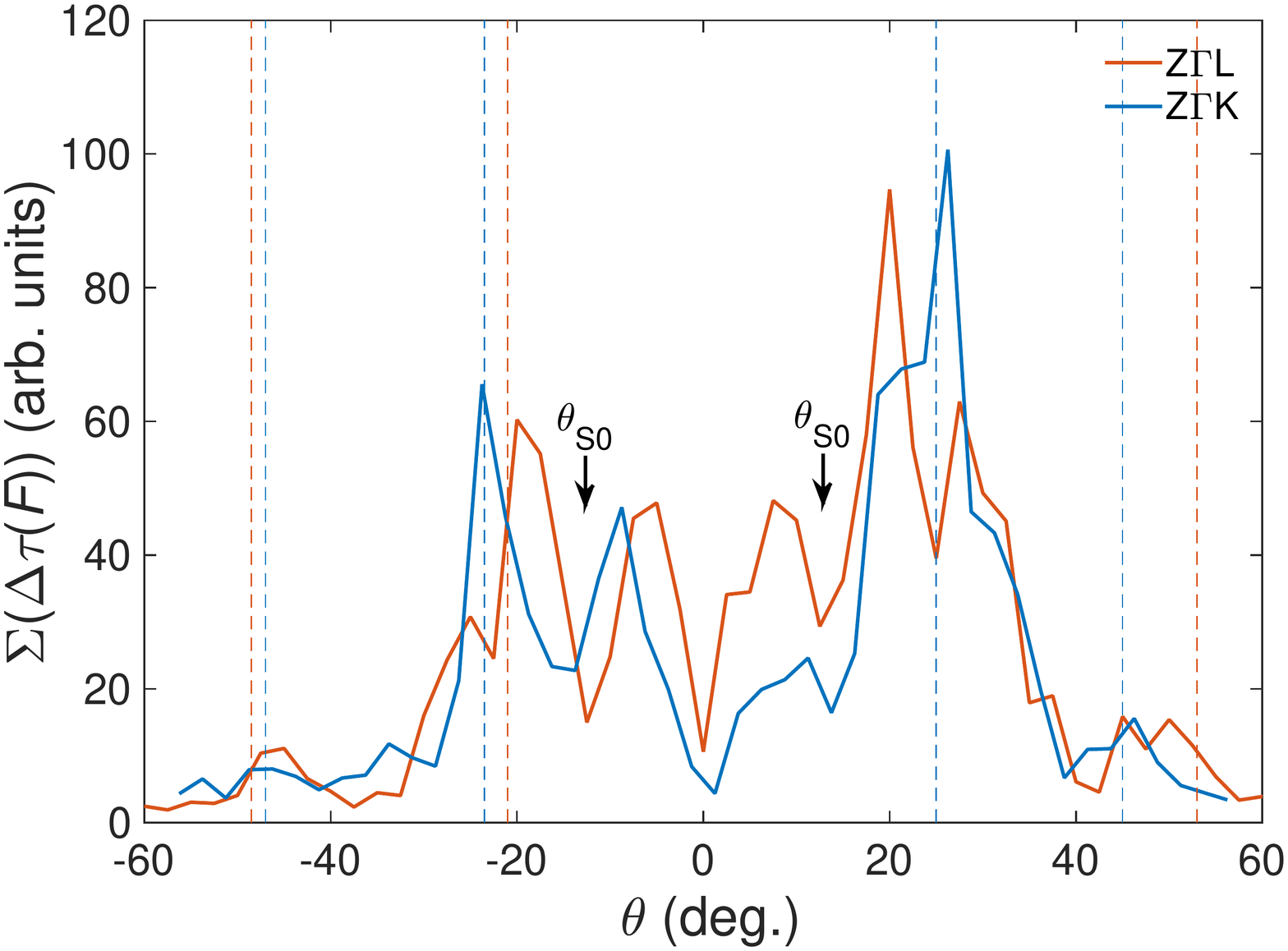}
	\caption{Angular dependence of the de Haas-van Alphen amplitude. The amplitude was calculated by discrete summation of the Fourier transforms Fig.~\ref{fig:FFTvTheta_Raw} for frequencies $F\geq25\,\mathrm{kT}$. Yamaji angles are indicated by vertical dashed lines. Note the absence of the first Yamaji angle in the $\mathrm{Z}\Gamma\mathrm{L}$ plane for positive angles.}
	\label{fig:AmplvTheta}
\end{figure}

\noindent Figure~\ref{fig:AmplvTheta} shows the angular dependence of the quantum oscillation amplitude, as determined by discrete summation of the Fourier transforms (Fig.~\ref{fig:FFTvTheta_Raw}) above $25\,\mathrm{kT}$. Generally the quantum oscillation amplitude is described by a superposition of the $\sin\{2\theta\}$ magnetic torque angular dependence of a 2D-electron gas, the Lifshitz-Kosevich reduction terms (temperature, Dingle and spin reduction term) \cite{Shoenberg,Lifshitz56,Dingle52} and in case of multiple extremal orbits, their interference \cite{Yamaji89}. Besides the strong peaking of the quantum oscillation amplitude at the Yamaji angles( $B\in \mathrm{Z}\Gamma\mathrm{K}:$ $\pm25^\circ$, $\pm46^\circ$; $B\in \mathrm{Z}\Gamma\mathrm{L}:$ $+21^\circ$, $+49^\circ$, $-53^\circ$)  (see also main article) and the spin reduction term, all other components follow a smooth angular dependence. Thus it is possible to evaluate the spin splitting i.e. the mean effective charge carrier moment or g-factor of a Fermi surface orbit by the observation of spin zeros. In Fig.~\ref{fig:AmplvTheta} we observe an unexplained suppression of the quantum oscillation amplitude at $13.5^\circ\pm1.0^\circ$. Due to warping of the Fermi surface and presence of two extremal orbit the amplitude is not fully suppressed. Following the spin reduction term:
\begin{eqnarray}
R_S=\cos\left\{\frac{\pi g m^*(\theta)}{2}\right\}\equiv0 \hspace{0.5em}\mathrm{with:}\hspace{0.5em} m^*(\theta)=\frac{m^*}{\cos\{\theta\}}
\end{eqnarray}
and taking into account the cyclotron masses of $m^*_\beta=(1.63\pm0.03) m_\mathrm{0}$ and $m^*_\alpha=(1.43\pm0.03) m_\mathrm{0}$ (see also main article), this angle corresponds to a $g$-factor of $g=(1.91\pm0.12)$. However, due to the periodicity of the spin reduction term, the determined $g$-factor is not unique and other solutions $g\in\{3.18;4.45;5.71;...\}$ might be possible \footnote[1]{Due to the large magnetic torque interaction experienced in our experiments, large amplitude oscillations are strongly damped, leading to a flattening of the angular dependence. In addition at large angles, which are necessary to distinguish between the higher order $g$-factors, the signal to noise is rather poor and the integral FFT amplitude suffers from a dominating noise floor.}.

\subsection{4. Density Functional Theory Calculations}

\begin{figure*}[tb]
\begin{minipage}{0.47\textwidth}
	\includegraphics[width=0.9\textwidth]{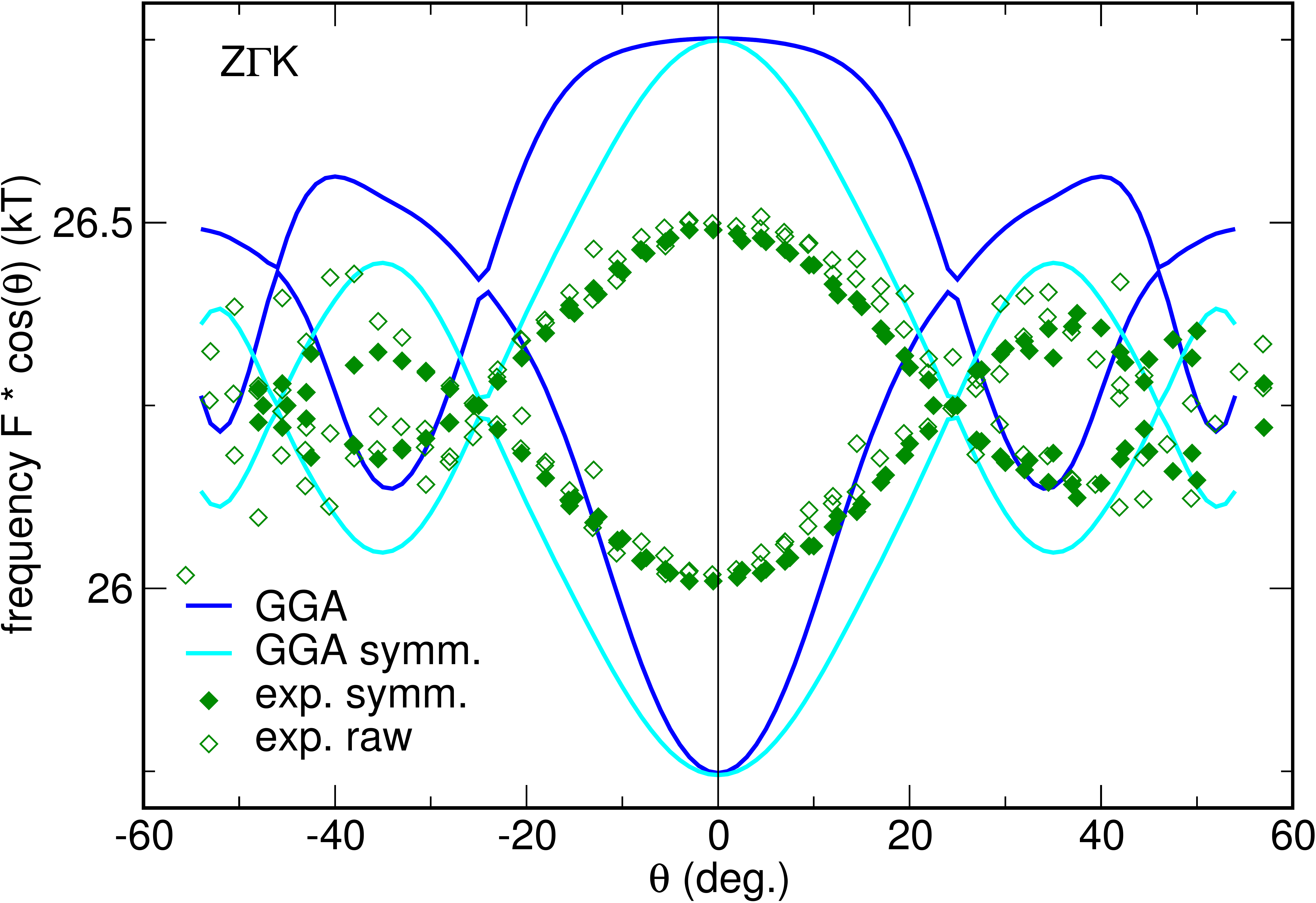}
\end{minipage}
\begin{minipage}{0.5\textwidth}
  \includegraphics[width=0.9\textwidth]{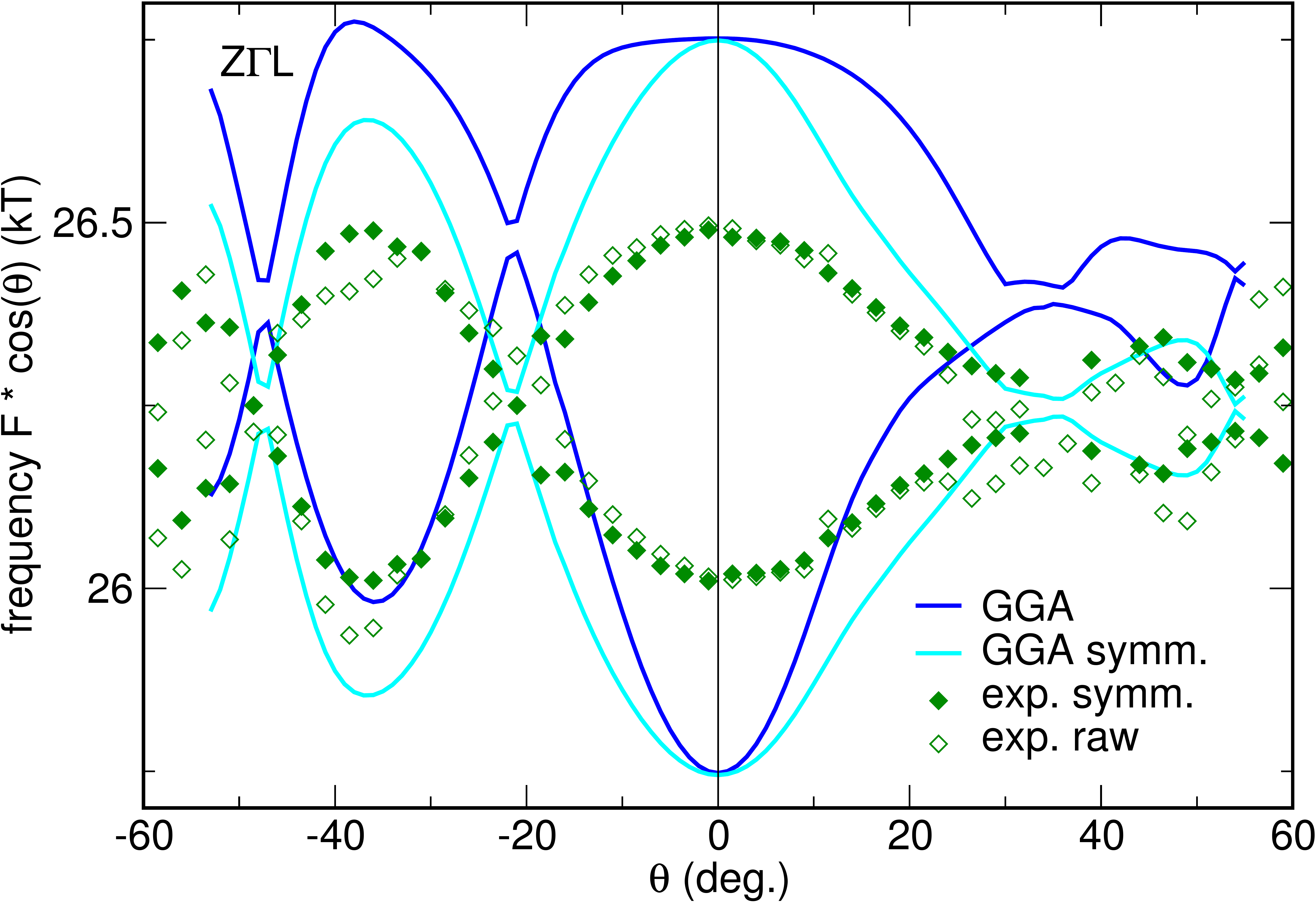}
\end{minipage}
\caption{Calculated dHvA cross sections applying GGA (dark blue). To facilitate the comparison with the experimental data (green symbols), the calculated frequencies have been shifted up by $\approx0.3\,\mathrm{kT}$. "Symmetrized" calculated data (see main text) are shown in light blue. The notation of field directions is provided in the main text.}
\label{fig:DFTCalcU0}
\end{figure*}

\begin{figure*}[tb]
\begin{minipage}{0.47\textwidth}
	\includegraphics[width=0.9\textwidth]{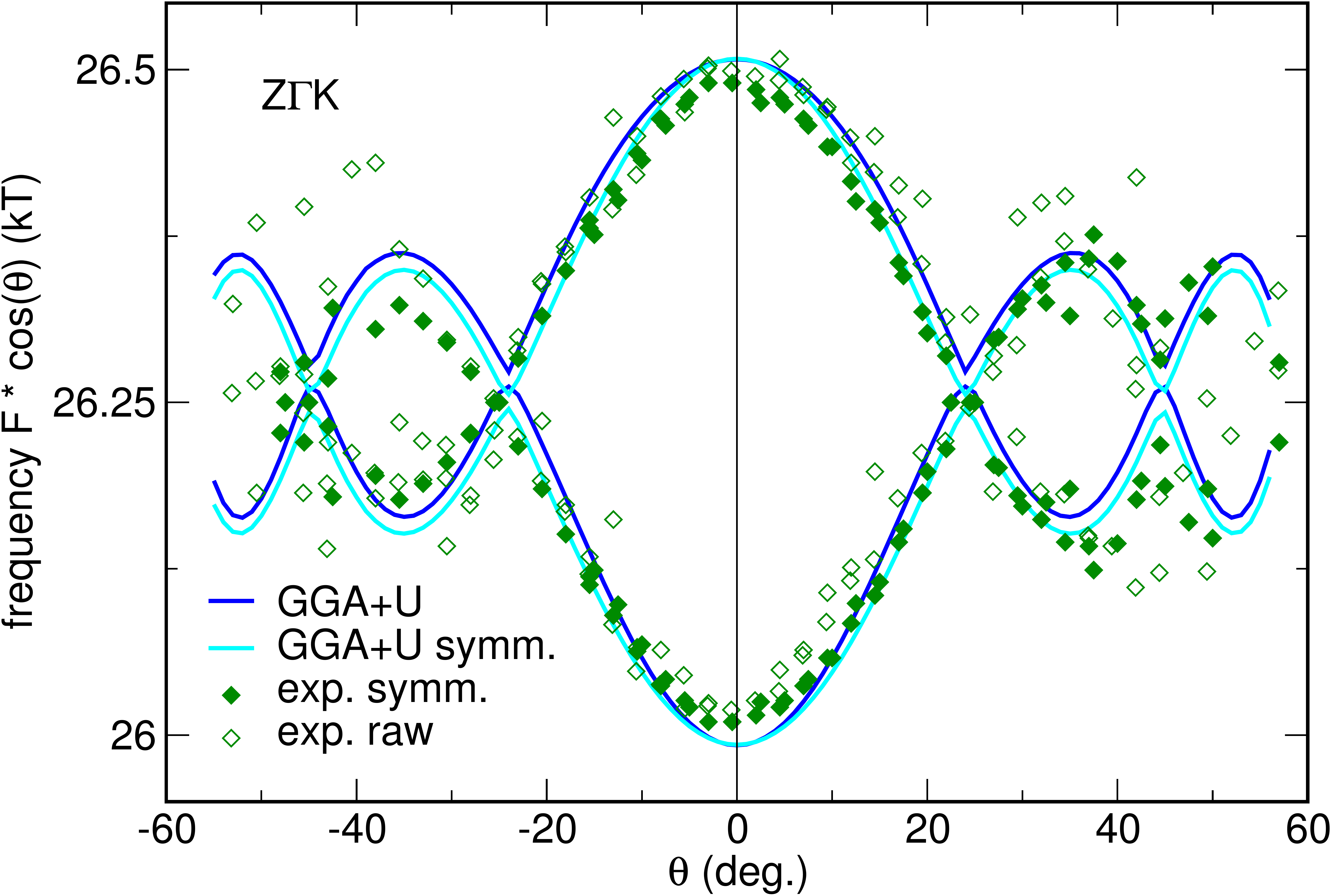}
\end{minipage}
\begin{minipage}{0.5\textwidth}
  \includegraphics[width=0.9\textwidth]{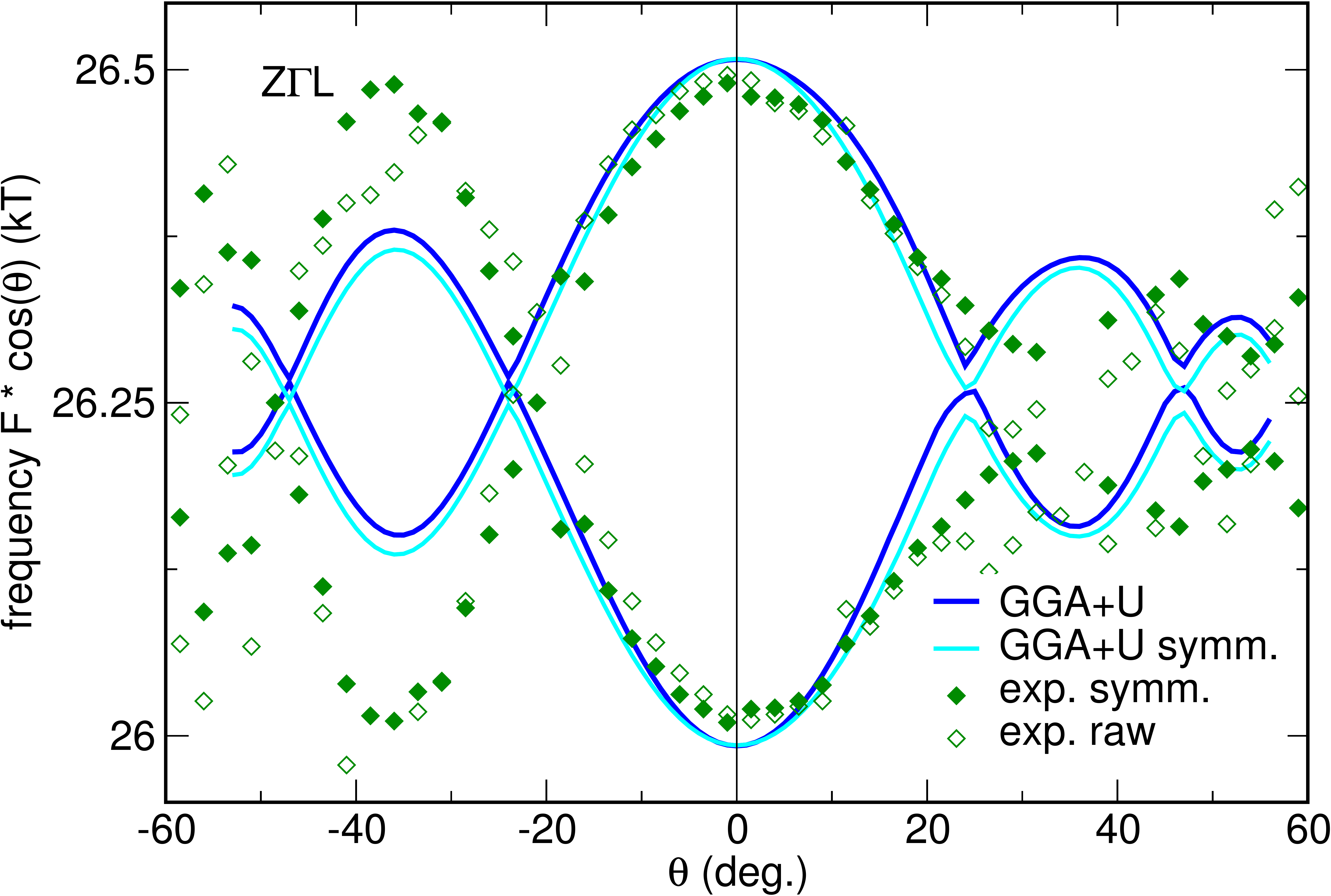}
\end{minipage}
\caption{Calculated dHvA cross sections applying GGA + $U$ (dark blue, $U=2.7\,\mathrm{eV}$, $J=0.5\,\mathrm{eV}$). To facilitate the comparison with the experimental data (green symbols), the calculated frequencies have been shifted up by $\approx0.2\,\mathrm{kT}$. "Symmetrized" calculated data (see main text) are shown in light blue. The notation of field directions is provided in the main text.}
\label{fig:DFTCalcU27}
\end{figure*}
Relativistic density functional (DFT) electronic structure calculations including spin-orbit coupling were performed on a $60\times 60\times60$ $k$-mesh, 18941 points in the irreducible wedge of the Brillouin zone. The spin-orbit (SO) coupling was treated non-perturbatively solving the four component Kohn-Sham-Dirac equation \cite{Eschrig}. For the exchange-correlation potential, within the general gradient approximation (GGA), the parametrization of Perdew-Burke-Ernzerhof \cite{Perdew96} was chosen. 

To obtain the rather small deviations from a purely 2D Fermi surface accurately, a self adjusting $k$-mesh was used to calculate the Fermi vectors. Interpolating the potential for the dense $k$-mesh of the self consistent calculation ($216.000$ $k$-points), the $k$-mesh was refined around the Fermi level iteratively to 1/32 of the original spacing, thus effectively covering approximately 2000 $k$-points in each direction of the Brillouin zone. The dHvA frequencies were evaluated on an angular mesh of $1^\circ$ and and 200 "slices" of the Brillouin zone along the respective field direction.

The calculated cross sections, compared with the experimental data, are shown in Fig.~\ref{fig:DFTCalcU0}. The calculated averaged frequency $F_0$ was sightly adjusted by ($0.3\,\mathrm{kT}$) to match the experimental value of $26.25\,\mathrm{kT}$, the deviation is likely caused by the difference in lattice parameters due to thermal expansion. Note that DFT calculations are based on the room temperature lattice parameters presented in \cite{Kushwaha17}. Quantum oscillation and ARPES data, however, are taken at $100\,\mathrm{mK}$ and $13\,\mathrm{K}$ respectively. For the GGA calculation, we obtain a good qualitative agreement with the experimental data with respect to shape and asymmetry of the Fermi surface. The dispersion along the $z$-direction, however, exceeds the experimental value by approximately a factor of two. The calculated Fermi velocities are sligthly underestimated, the corresponding bare band masses are somewhat overestimated.

Simulating the Coulomb correlation in the Rh-$4d$ shell in a mean field way, applying the GGA + $U$ scheme ($U=2.7\,\mathrm{eV}$, $J=0.5\,\mathrm{eV}$), the overall agreement with the experimental data is improved (see Fig.~\ref{fig:DFTCalcU27}). For the applied value $U=2.7\,\mathrm{eV}$, the dispersion along $z$ agrees well with the experiment. In contrast, the asymmetry of the calculated FS (see Fig.~\ref{fig:DFTCalcU27} right panel - Z$\Gamma$L) is underestimated by the GGA + $U$ scheme. Compared with the pure GGA calculations, however, the calculated Fermi velocities and the corresponding bare band masses are significantly improved with respect to the experimental data.

\end{document}